\documentclass[%
 reprint,
 superscriptaddress,
 amsmath,amssymb,
 aps,
 multiline,
prb,
]{revtex4-2}

\usepackage{amsmath}
\usepackage{float}
\usepackage{amssymb}
\usepackage{graphicx}
\usepackage[%
  colorlinks=true,
  urlcolor=blue,
  linkcolor=blue,
  citecolor=blue
]{hyperref}

\renewcommand\vec{\mathbf}

\newcommand{\rr}{\vec{r}}

\def \equi#1{\mathrel{\mathop{\kern 0pt\sim}\limits_{#1}}}

\begin{document}
	\title{
    %Two-dimensional melting scenarios and tricriticality in a dilute? clock model
    An on-lattice model for two-dimensional melting 
    }
    
	\author{Antonin Rogé}
    \affiliation{ENS de Lyon, CNRS, LPENSL, UMR5672, 69342, Lyon cedex 07, France}
    
	\author{Peter Holdsworth}
    \affiliation{ENS de Lyon, CNRS, LPENSL, UMR5672, 69342, Lyon cedex 07, France}
    
	\author{Alexis Poncet}
    \affiliation{ENS de Lyon, CNRS, LPENSL, UMR5672, 69342, Lyon cedex 07, France}
	
    \begin{abstract}
We introduce an on-lattice minimal model of two-dimensional melting, a clock model with vacancies with control parameters temperature, chemical potential and external field. We study the full phase diagram using cluster Monte Carlo, Tensor Network Renormalisation Group and self-consistent mean field methods. A non-universal ensemble of melting scenarios is observed including the KTHNY scenario of two Berezinskii-Kosterlitz-Thouless phase transitions delimiting a quasi-long-range ordered phase. The upper transition evolves into first-order transition in analogy with simulations on hard and soft discs and at low chemical potential there is a single transition between ordered high density and disordered low density phases. The evolution is via a special tricritical point that is accessed in detail. 
    This work therefore provides a 
    framework for understanding the crossover between 2D melting scenarios and offers insights into the behavior of realistic systems.
    %\todo[inline]{Vérifier qu'on est contents de l'abstract}
    \end{abstract}

	\maketitle

    \section{Introduction}
    Two-dimensional melting is a paradigmatic problem of statistical mechanics. 
    Following the seminal works of Berezinskii~\cite{Berezinskii1971,Berezinskii1972} and Kosterlitz and Thouless~\cite{Kosterlitz1973} (BKT), Halperin and Nelson predicted~\cite{Halperin1978,Nelson1979} a two-stage melting scenario in which two BKT transitions delimit a hexatic liquid-crystal phase exhibiting quasi-long-range orientational order. This exotic phase is sandwiched between a solid phase characterized by Young~\cite{Young1979} and a fluid phase. The KTHNY scenario, although non-universal could occupy an extensive part of the pressure-temperature phase diagram \cite{Nelson1979}. Other scenarios could include a single first-order transition along low or high pressure isobars, as in three-dimensional melting or a mixture of first-order and BKT transitions. The fluid could also show a liquid-gas transition if attractive interactions are present. Intense experimental~\cite{Murray1987,Zanh_1999,Peng2010,Schockmel2013,Deutschlander_2013} and numerical studies~\cite{Chen1995,Prestipino2011,Durand2019} have provided evidence of KTHNY melting and subsequent work does indeed suggest a wide range of different scenarios depending strongly on the microscopic details of the interactions~\cite{Saito1982,Chui1983,Domany1984,Strandburg1988,Marcus1996,VanEnter2002}. These include a single discontinuous transition for specific core-softened potentials~\cite{Prestipino2012,Dudalov2014,Du2017}.

    Advances in numerical techniques and computing power have opened the door to a more systematic approach to simulations of simple fluids~\cite{Bernard_2011,Engel2013,Kapfer_2015,Qi2014,Hajibabaei2019,Tsiok2022}. KTHNY melting was confirmed for discs with soft, repulsive power-law interactions, but as the interaction becomes harder  the fluid-hexatic transition becomes first-order~\cite{Bernard_2011,Engel2013,Kapfer_2015,Qi2014,Hajibabaei2019} leaving a single BKT transition into the solid phase. Surprisingly, this scenario is maintained for hard discs. Discs interacting via a Yukawa potential can be shown to exhibit either of these scenarios depending on the screening length~\cite{Kapfer_2015}. However, characterizing what happens in-between these two scenarios, the nature of the point at which the second transition switches from continuous to first-order, or mapping out complete phase diagrams for more complex fluids remains a daunting task.
    %for continuous space systems.
    
    In this article, we propose a minimal on-lattice model for two-dimensional melting that exhibits a full range of melting scenarios covering a complete phase diagram. Its simplicity allows us to investigate the details of the transition lines, as well as the special tricritical point connecting BKT and first-order lines. The model could be further extended to include a liquid-gas phase transition.  
    
    We study the site diluted $p-$state clock model, Eq.~\eqref{eq:H}. %presented in detail in the next section. 
    Predating KTHNY, José {\it et al.} \cite{Jose1977} showed that an XY-spin model with a $p-$fold crystal field perturbation with $p>4$ presents a scenario with two BKT phase transitions
    as the temperature $T$ is reduced. 
    Reminiscent of two-stage melting, the intermediate phase between the two BKT transitions has emergent continuous symmetry and quasi-long-range orientational order as in the hexatic liquid crystal phase. True long range orientational order is established in the low temperature phase with the breaking of the $Z_p$ symmetry. Continuous symmetry emerges over a finite temperature range even in the discrete $p-$state clock model~\cite{Elitzur1979,Lapilli2006,Li2022} and we have taken advantage of this for %its
    numerical efficiency. Including site dilution via a chemical potential, $\mu$ introduces density fluctuations, as in Blume-Capel type models~\cite{Blume1966,Capel1966,Blume1971} introduced to study helium-3--helium-4 mixtures. Here, 
    a first-order line gives way to a continuous transition via a tricritical point, $(T_t,\mu_t)$~\cite{Kwak2015,Zierenberg2017,Moueddene2024}. An analogous tricritical point occurs in vector models separating a first-order line from a BKT transition~\cite{Berker1979,Cardy1979,Dillon2010,Santos2018,Skovdal2023}. Recently, there has been a renewed interest in such models in the context of conformal field theory~\cite{Li2022,Guo2024,Homma2024} and we have built on this work to calculate the central charge and critical exponents using tensor-network renormalization group methods.
    
\begin{figure}
    	\centering
    	\includegraphics[scale = 1]{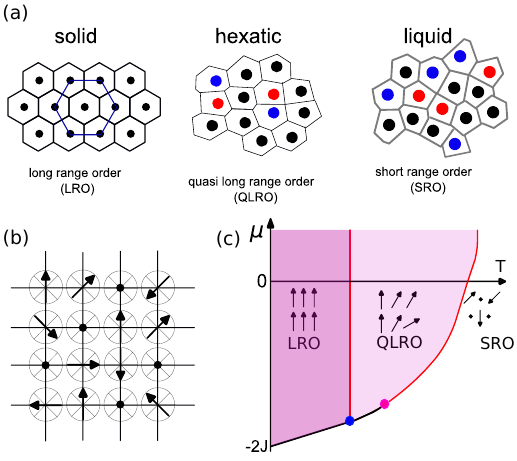}
    	\caption{\textbf{Analog of 2D melting in a spin model.}
    		(a) The three phase of 2D crystals and their orientational order.
    		(b) The lattice model, see Eq.~\eqref{eq:H}. 
    		The spins can take $p=8$ discrete orientations $\theta_i = 2k_i\pi/p$ with integer $k_i$ and have two possible lengths $\eta_i=0$ (dots) or $\eta_i=1$ (arrows).
    		(c) Qualitative phase diagram in the $(T, \mu)$ plane showing the same types of order as 2D solids: long-range order (LRO), quasi-long-range order (QLRO) and short-range order (SRO). The red lines denote continuous BKT transitions while the black line is a first-order transition.
    		Also shown are the tricritical point where the high-temperature BKT transition becomes first-order (magenta), and the point at which the low-temperature BKT transition meets the first-order line (blue). See Fig.~\ref{fig:phaseDiag} for a quantitative phase diagram. We also highlight that the magenta area is a plane of critical points with QLRO order.
        }
    	\label{fig:clock}
    \end{figure}

    A qualitative $\mu-T$ phase diagram for the model and the 
    analogy with two-dimensional melting are illustrated in Fig.~\ref{fig:clock}. For large $\mu$ the density remains high through the orientational ordering which occurs via two BKT transitions, as for the $p-$fold field problem. On lowering $\mu$ the higher temperature transition becomes first-order via a tricritical point and the ordering is accompanied by a density jump. As $\mu$ is lowered further the ordering transitions merge, leaving a single first-order transition between a low density, disordered lattice fluid phase and high density ordered phase. The model therefore offers iso-potentials with three of the possible ordering scenarios encountered in simulations of fluids and in experiments: two BKT transitions, a first-order followed by a BKT transition and a single first-order transition for low values of $\mu$. 
    
    The article is organized as follows.
    In section~\ref{s:model}, we introduce the clock model with vacancies and 
    present the phase diagram calculated from a self-consistent mean-field theory capturing the density fluctuations. 
    To go beyond this approximation, we introduce Monte-Carlo methods, with both local moves and cluster updates~\cite{Swendsen1987,Wolff1989,Zierenberg2017} as well as tensor network renormalization techniques~\cite{Levin2007,Gu2009,Yang2017,Guo2024,Homma2024}, used to 
    locate the critical lines and obtain quantitative values for the critical exponents. 
    In section~\ref{s:scenarios} the three possible melting scenarios are analyzed. 
    The quantitative phase diagram is presented in section~\ref{s:tricritical}, detailing
    the evolution of the high-temperature BKT transition with decreasing $\mu$. The tricritical point is located and the tricritical exponents evaluated. The field scaling dimension is found to be consistent with the Ising tricritical exponent but that associated with density fluctuations appears to be quite different. Finally we extend the phase diagram to include an external field $h$ coupling to the orientational degrees of freedom.
    In section~\ref{s:discussion}, we discuss our findings in the context of 2D melting before concluding with some perspectives.

    \section{Clock model with vacancies} \label{s:model}
    
    \subsection{The model}
    The site diluted  $p$-state clock model on the square lattice of  $N=L^2$ sites with linear size $L$ has generalized Hamiltonian 
    \begin{equation} \label{eq:H}
    \mathcal{H} = -J \sum_{\langle i,j \rangle} \eta_i \eta_j \cos(\theta_i - \theta_j) - \mu \sum_i \eta_i -h\sum_i \eta_i\cos{(\theta_i)}.
    \end{equation}
    Site $i$ is occupied by a rotor, or spin $\vec S_i$ with discrete orientation $\theta_i = \frac{2k_i\pi}{p}$ with $k_i$ an integer taking values $0, \dots, p-1$. The spin has length $\eta_i \in \{\epsilon, 1\}$ in the limit $\epsilon \rightarrow 0$. The energy scale for fixing a spin of unit length is set by the chemical potential $\mu$. This is an excess term, measured with respect to a reference state, the non-interacting lattice fluid with hard core repulsion. 
    Neighboring spins interact in the Hamiltonian with a ferromagnetic coupling constant $J=1$. The spins may also couple to an external field $\vec h=h\hat{\vec x}$ that breaks the symmetry along $\theta_i=0$. The sum over nearest neighbors is designated 
    $\langle i,j \rangle$ and periodic boundaries are used throughout. As the hexatic phase is built around six-fold coordination, the natural model to use would appear to be $p=6$. However, in order to avoid issues specific to the discrete clock model \cite{Lapilli2006} we 
    focus on the case of $p=8$ which, in the absence of vacancies orders via two well documented BKT transitions.

    The interaction term in Eq.~\eqref{eq:H} has discrete translational symmetry and this is not reduced in the low temperature phases so that, unlike a real fluid, translational and orientational order are decoupled. 
   Retaining a phantom spin of vanishing length at an empty site ensures that the density of spins of unit length approaches $\frac{1}{2}$ at high temperature rather than $\frac{p}{p+1}$, consistently with site dilution in classical spin models with continuous symmetry~\cite{Berker1979}.

\begin{figure}
	\centering
	
	\includegraphics[width=\columnwidth]{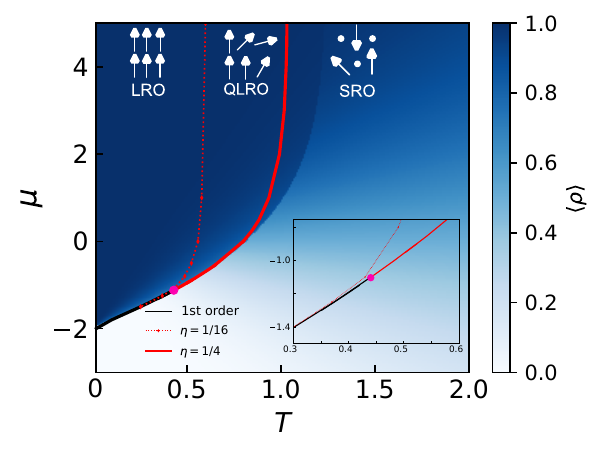}
	\caption{\textbf{Variational phase diagram.} 
		Average density $\langle\rho\rangle_\mathrm{MF}$ from the variational Hamiltonian, Eq.~\eqref{eq:approx_mf}. The BKT transition corresponding to $\eta=1/4$ (full red line) occurs before the discontinuity at high values of $\mu$. On the left of the line $\eta=1/16$ (dotted red line) the system is unstable to $p$-fold symmetry breaking implying long-range order.
	}
	\label{fig:HartreeFock}
\end{figure}

\begin{figure*}
	\centering
	\includegraphics[scale = 1]{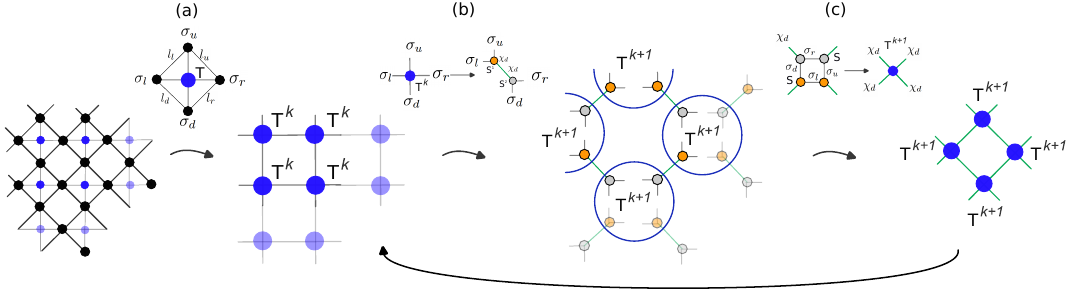}
	\caption{\textbf{Tensor network renormalization group.}
		(a) Starting from the initial lattice (Fig.~\ref{fig:clock}b) in black, we define the local rank-4 tensor $T$ in blue on the dual lattice.
		(b) This tensor %can be 
        is decomposed into two rank three tensors by a SVD.
		(c) Neighboring three-leg tensors 
        %can then be 
        are then recombined by groups of four, leading to a coarse grained lattice with half of the number of initial tensors (rotated by 45 degrees). The TNR method consists of iterating this procedure until a single tensor remains.}
	\label{fig:tnr}
\end{figure*}

   \subsection{Hartree-Fock approximation}
    An approximate
    phase diagram can be generated using a self-consistent mean field technique in the spirit of the 
    Hartree-Fock approximation for the XY model~\cite{Portelli2001}. A mean field Hamiltonian (with $h=0$) can be written   
    \begin{equation}
        \mathcal{H}_\mathrm{MF} = \mathcal{H}_0 + \frac{J_\mathrm{eff}}{2} \sum_{\langle i,j \rangle} (\theta_i - \theta_j)^2 - \mu_\mathrm{eff}\sum_i \eta_i,
        \label{eq:approx_mf}
    \end{equation}
    with self-consistent parameters $J_\mathrm{eff}$ and $\mu_\mathrm{eff}$. As in the case of the $XY$ model, the anharmonic contributions to the cosine interactions can be renormalised into an effective harmonic coupling. In addition, site dilution is also taken into account at the mean field level. 
    The $\{\eta_i, \theta_i\}$ variables are decoupled and $\theta_i$ is taken to be a continuous variable.
    The Feynman-Bogolyubov variational principle~\cite{Feynman2018} is used to obtain self-consistent equations for $J_\mathrm{eff}$ and $\mu_\mathrm{eff}$. The free energy $\Omega$ 
     %$= -\frac{1}{\beta}\ln \mathcal{Z}$ where $\mathcal{Z}$ is the partition function 
     is approximated from above as
    \begin{equation}
        \Omega \leq \Omega_\mathrm{MF} + \langle \mathcal{H} - \mathcal{H}_\mathrm{MF}\rangle_\mathrm{MF},
    \end{equation}
    where the average %$\langle \rangle_\mathcal{MF}$
    is taken over the distribution generated by Eq.~\eqref{eq:approx_mf}. 
   Performing the minimization 
   leads to coupled equations for $ (J_\mathrm{eff}, \mu_\mathrm{eff})$,
    \begin{equation}
        \begin{cases}
       J_\mathrm{eff} = J \langle\rho\rangle_\mathrm{MF}^2 e^{-T/4J_\mathrm{eff}} ,\\
        \mu_\mathrm{eff} = 
        \mu + 2J \langle\rho\rangle_\mathrm{MF} e^{-T/4J_\mathrm{eff}}.
        %\mu + \frac{2J}{1 + e^{-\beta \mu_\mathrm{eff}}} e^{-T/4J_\mathrm{eff}}.
        \end{cases}
    \label{eq:system_eq}
    \end{equation}
    The average density is 
        $\langle \rho \rangle_\mathrm{MF} = -\frac{\partial \Omega_\mathrm{MF}}{\partial \mu_\mathrm{eff}} = [1 + e^{-\beta \mu_\mathrm{eff}}]^{-1}$.
    
    The set of equations can be solved 
    numerically.
A discontinuity in the density occurs along a first-order line in the $(\mu,T)$ plane as illustrated in Fig.~\ref{fig:HartreeFock}. However, the first-order transition should be preempted by a BKT transition if the anomalous exponent, 
$\eta = \frac{T}{2 \pi J_\mathrm{eff}}$, 
exceeds the threshold $\eta=1/4$ before the predicted discontinuity. At this point, the spin stiffness takes on the universal value 
$K_\mathrm{eff}=\frac{J_\mathrm{eff}}{T}=\frac{2}{\pi}$. 
The renormalization leads to a transition temperature, $T_{KT}=0.89J$ in the $XY$ model \cite{Minnhagen1987,Gupta1988} and $1.35J$ within the Villain approximation \cite{Villain1975} and the Hartree-Foch method predicts these values with around $10\%$ error. Heating beyond it, 
    vortices unbind, $K_\mathrm{eff}$ jumps to zero, and the variational approach fails~\cite{Portelli2001}. 
    The line $\eta=1/4$ is reported in Fig.~\ref{fig:HartreeFock}. 
This threshold intercepts the first-order line at a  tricritical point estimated to be
$\tilde{T}_t\approx 0.45, \tilde{\mu}_t \approx -1.15$, in qualitative agreement with the numerical estimate shown below. 

An estimate of the point of interception of the first-order transition with the low temperature BKT transition can also be made. Through a duality transformation Jos\'e {\it et. al.} \cite{Jose1977} were able to show that the spin wave Hamiltonian Eq.~\eqref{eq:approx_mf} becomes  unstable to $p-$fold symmetry breaking, at a threshold $K_\mathrm{eff} = p^2/(8\pi)$~\cite{Jose1977}, or $\eta=1/16$ for $p=8$. We find an intercept at $\tilde{T}_a=0.35, \tilde{\mu}_a\approx -1.31$, again in qualitative agreement with our numerical estimate. Note that, neglecting the renormalization of the coupling constant, one finds a zeroth order approximation for the transition 
$T_1^{\ell}=\frac{\pi}{8}J=0.39J$, in reasonable agreement with numerical estimates throughout the phase diagram.

    Fig.~\ref{fig:HartreeFock} shows the variational phase diagram. The finite slope of the phase boundary as $T\rightarrow 0$ 
    is a direct result of the residual entropy of the vacuum~\cite{Berker1979}. In the Blume-Capel model this initial slope is zero.
    The slope can be estimated using the Clapeyron equation
    \begin{equation}
        \frac{d\mu}{dT}=-\left(\frac{S_f-S_s}{N_f-N_s}\right),\label{Clapeyron}
    \end{equation}
    where $s$ and $f$ indicate the solid (ordered) and fluid (disordered) phases, $S_x$ and $N_x$ are the entropy and number of spins of unit length in phase $x$. At the zero temperature phase boundary, $N_s=N, S_s=0$ and $N_f=0,S_f=Nk_B\ln(p)$, giving a slope of $2.079k_B$ for $p=8$. 
    %\textcolor{red}{ From numerical differentiation along the phase boundary we find a slope of 2.1103$k_B$ in excellent agreement with this prediction}. 
    Despite providing a qualitative phase diagram, our variational method
    fails completely in all aspects of calculations for the QLRO phase, the BKT transitions and the tricritical point. For this we turn to numerical methods.

    \subsection{Monte-Carlo simulations}

    \label{sec:num_approaches}

    The Monte-Carlo method (MC) corresponds to a mixed Metropolis algorithm~\cite{Zierenberg2017} with two kinds of moves: local updates for the lengths $\eta_i$ and global cluster updates for the orientations $\theta_i$~\cite{Wolff1989}. 
    We used linear size $L=16$ to $128$ ($N=L^2=256$ to $16384$).
    The system was equilibrated by a run of $200\,000$ MC steps, where a single MC step consists of 1 cluster update and 100 metropolis updates per spin. Once thermalization was reached, $100\,000$ sampling steps were performed, each separated by 200 MC steps.
    
    The magnetization $\vec m$ and the density $\rho$ of a given configuration are
    \begin{equation}
    	\label{eq:m_rho}
    	\vec m=\frac{1}{N}\sum_i \vec{S}_i, \qquad \rho=\frac{1}{N}\sum_i \eta_i. 
    \end{equation}
    In zero field and unless otherwise stated the magnetization was taken as the average of the modulus $\langle m\rangle=\langle |\vec m|\rangle$. In finite field we used the average projection along the field axis $\langle m\rangle=\langle \vec m\rangle . \hat{x}$.
    
    Calculated quantities include average values $\langle m\rangle$ and $\langle\rho\rangle$,  the magnetic and density susceptibilities
    \begin{align}
    	\label{eq:chi_mag}
    	\chi_{\mathrm{mag}, L}&=\beta L^2[\langle m^2\rangle - \langle m\rangle^2], \\
    	\label{eq:chi_rho}
    	\chi_{\rho, L}&=\beta L^2[\langle \rho^2\rangle - \langle \rho\rangle^2],
    \end{align} 
    the latter being related to the compressibility  $\kappa= \frac{1}{\langle \rho\rangle^2}\chi_{\rho, L}$,
    and the Binder cumulant
    \begin{equation}
    	U_L = 1 - \frac{\langle m^4\rangle}{3\langle m^2\rangle^2}  \label{eq:UL},
    \end{equation}
    which exhibits scale invariance in the QLRO phase~\cite{Lapilli2006,Hasenbusch_2008}.
    Finally, to locate the low-temperature transition of the clock model, we use an orientational Binder cumulant~\cite{binder_cumulants_clock}
    \begin{align}
    	U_\phi &= 1 - \frac{\langle m_\phi^4\rangle}{2\langle m_\phi^2\rangle^2}, \label{eq:Uphi}
    \end{align}
    where $m_\phi = \cos(p\phi)$ is the
    the $p-$fold order parameter, with $\phi = \arctan(\frac{\sigma_y}{\sigma_x}), \sigma_x = \sum_i \cos(\theta_i), \sigma_y = \sum_i \sin(\theta_i)$.

     \subsection{Tensor network renormalization}
     
     The MC approach suffers from strong hysteretic effects around the first-order transition line, preventing one to locate it precisely.
     %and is ineffective in precisely locating the first-order line in the phase diagram.
     %Given these limitations 
     We therefore
     supplement the MC simulations with tensor network renormalization (TNR) calculations, the main ideas of which are resumed below, see Refs.~\cite{Levin2007,Gu2009,Yang2017,Guo2024,Homma2024} and App.~\ref{app:tnr} for details.
     The starting point is to write the partition function $\mathcal{Z}$ as a product of local tensors $T^k$ on the nodes $k$ of the dual lattice, see Fig~\ref{fig:tnr},
   \begin{equation}
    \mathcal{Z} = \sum_{\{\eta_i, \theta_i\}} e^{-\beta \mathcal{H}( \{\eta_i, \theta_i\})} = \mathrm{tTr} \left( \bigotimes_{k=1}^{N/2} T^k_{u,r,d,l} \right),
    \end{equation}
where the sum $\{\eta_i, \theta_i\}$ runs over all the configurations (lengths and angles) of the $N$ spins, $\beta$ is the inverse temperature, the Hamiltonian $\mathcal{H}$ is defined in Eq.~\eqref{eq:H}, and $\mathrm{tTr}$ is the tensorial trace of the product of tensors. The local tensor $T^k$ is defined at the center of a plaquette of four spins denoted by $\{u,r,d,l\}$,  see Fig~\ref{fig:tnr}. It accounts for the Boltzmann weight of interaction between neighboring spins
and reads 

\begin{equation}
	T^k_{u,r,d,l} = \prod_{i\in \{u,r,d,l\}} e^{\frac{\beta}{2}\eta_i\left(\mu + h\cos\theta_i\right) }  \prod_{\langle i,j\rangle} 
	e^{\frac{\beta}{2} \eta_i\eta_j \cos(\theta_i-\theta_j)},
	\label{eq:local_tensor}
\end{equation}
where the product $\langle i,j\rangle$ runs over the four links in the plaquette.
For $p$ orientational degrees of freedom, $T^k$ is a rank-4 local tensor of depth $2p$. Thanks to the translational symmetry and nearest neighbor interactions, all the initial local tensors $T^k$ are identical. 

At each iteration the tensor network is coarse-grained, dividing the number of tensors by two, up to the point where there is a unique tensor that can be traced to obtain the partition function.
%To avoid exponential growth of the tensor size, 
Singular value decomposition (SVD) is used at each step to decompose a $T$ tensor into two $S$ tensors of rank 3,
\begin{equation}
    T_{u,r,d,l} = \sum_{\alpha} S^1_{u,r,\alpha} S^2_{\alpha,d,l},
\end{equation}
where the dimension of the indice $\alpha$ is at most the so-called bond dimension $\chi_d$. 
We decompose the tensors $T$ geometrically in two ways by grouping different indices to construct the $S$ tensors: $ T \rightarrow (S^1_{u,r,\alpha}, S^2_{\alpha, d, l})$ or  $T \rightarrow (S^1_{u,l,\alpha}, S^2_{\alpha, r, d})$ and obtain an octagonal configuration, see Fig.~\ref{fig:tnr}. The
%coarse grain
four rank-3 $S$ tensors are then combined together to get a new local rank-4 tensor of dimension $(\chi_d, \chi_d, \chi_d, \chi_d)$.
Having $\chi_d \rightarrow \infty$ would lead to a loss-free procedure.
In our study, we mostly used $\chi_d=40$, and up to $\chi_d = 64$ around the tricritical point. 

Several local optimization methods have been proposed to reduce the systematic error induced by truncation~\cite{Yang2017,Homma2024}.
In our study we use the nuclear-norm regularization for Tensor Network Renormalization (NNR-TNR) algorithm~\cite{Homma2024} which has recently been shown to be efficient for the extraction of a stable fixed point tensor for 2D classical systems on lattice. From this procedure we extract the partition function and the first derivatives of the free energy (magnetization $\langle m\rangle$ and density $\langle\rho\rangle$) using the %usual
impurity tensor renormalization method ~\cite{higher_order_moments, phys_obs}, see also App.~\ref{app:tnr}.

 TNR algorithms are also well suited for extracting field-theoretic quantities.
 A very useful one is the gauge invariant~\cite{Li2022},
 \begin{equation}
 	\chi_\mathrm{inv} = \frac{\left( \sum_{ij} T_{ijij} \right)^2}{\sum_{ijkl} T_{ijkl} T_{klij}}, \label{eq:chi_inv}
 \end{equation}
 with $T$ the final tensor of the TNR. This quantity  is directly linked to the symmetry of the system, see App.~\ref{app:tnr}.
 One has $\chi_\mathrm{inv} =1$ for a single disordered phase, $\chi_\mathrm{inv}=p$ when $p$ phases coexist, while at a critical point, $\chi_\mathrm{inv}$ takes a non-trivial universal value.
 As we show below,
$\chi_\mathrm{inv}$ can also be used to locate the first-order transition line.
 In addition to the gauge invariant, all quantities from conformal field theory (CFT) can be calculated from the final tensor, in particular the central charge $c$ and the scaling dimensions $\Delta_\alpha$~\cite{Li2022,Guo2024,Homma2024}, see App.~\ref{app:tnr}.
 The renormalization group eigenvalue exponents $y_\alpha$ are then retrieved from the scaling dimensions $\Delta_\alpha$ as $y_\alpha = d - \Delta_\alpha$ with $d=2$ the space dimensionality of the system.

  \begin{figure}
  	\centering
  	\includegraphics[scale = 1]{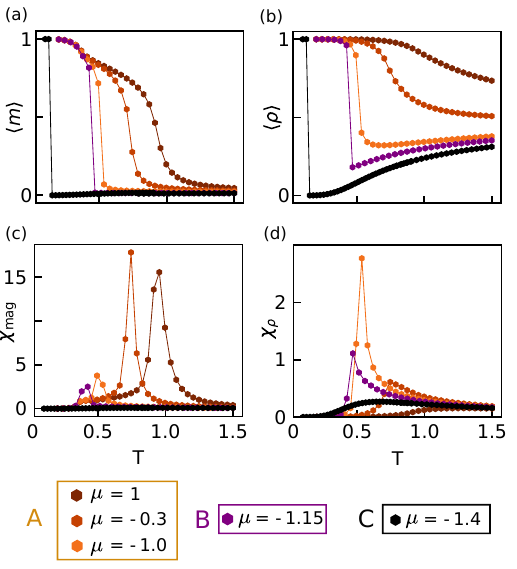}
  	\caption{\textbf{Three melting scenarios triggered by decreasing $\mu$, from MC simulations.} ($L = 48$)
  		(a) $\langle m\rangle$ as a function of $T$ for different $\mu$. Decreasing $\mu$ below $-1.15$ leads to a discontinuity.  
        (b) $\langle \rho\rangle$ as a function of $T$ showing a discontinuity for $\mu < -1.15$. %This highlights a first-order transition associated with density fluctuations. 
  		(c) $\chi_\mathrm{mag}$ {\it vs. } $T$. The peak at $T_2$ is reduced in amplitude as $\mu$ decreases.
        %showing that the criticality is less impacted by magnetic features.
  		(d) $\chi_\rho$ {\it vs.} $T$. 
        %Contrary to $\chi_\mathrm{mag}$, the density fluctuations $\chi_\rho$ increase when decreasing $\mu$, proving that density becomes relevant.
        Data shows development of a peak for intermediates of $\mu$, characteristic of a tricritical point.
        %\todo[inline]{Pas entièrement satisfait c/d}
  	}
  	\label{fig:scenarios}
  \end{figure}

    %%%%%%%
    \section{Three melting scenarios} \label{s:scenarios}

The three melting scenarios discussed in Fig.~\ref{fig:clock} can be identified.  This is illustrated in Fig.~\ref{fig:scenarios} where Monte-Carlo data
 for $\langle m\rangle$, $\langle \rho\rangle$, $\chi_{\mathrm{mag},L}$ and $\chi_{\rho,L}$,  Eqs.~\eqref{eq:m_rho}-\eqref{eq:chi_rho}, is plotted
 along different isopotentials for a system of size $L=48$.

    Scenario $A$ corresponds to the highest values of $\mu$ ($\mu=1,-0.3,-1$). It features a continuously increasing magnetization with decreasing temperature with two inflexion points, see Fig.~\ref{fig:scenarios}a. This is characteristic of BKT transitions when observed in finite size~\cite{Bramwell1993,Bramwell1994,Lapilli2006,Skovdal2023}.
   The high-temperature transition is characterized by a peak in the magnetic susceptibility~\cite{Bramwell1993,Bramwell1994}, see Fig.~\ref{fig:scenarios}c. The peak disappears as $\mu$ decreases towards the tricritical point that marks the emergence of a discontinuous transition. The approach of the tricritical point is also characterized by increasing density fluctuations, Fig.~\ref{fig:scenarios}d.

    Scenario $B$ corresponds to a first-order transtion, and then a BKT transition on reducing the temperature. It occurs for the chemical potential between $\mu_{t}\approx -1.1137$ and $\mu_a\approx -1.22$.
    Fig.~\ref{fig:scenarios} shows an isopotential with $\mu=-1.15$ where discontinuities in both $\langle m\rangle$ and $\langle\rho\rangle$ are observed at the higher temperature transition accompanied by a small peak in $\chi_{\mathrm{mag}, L}$ and a cusp in $\chi_{\rho, L}$. An inflection in the magnetisation and second peak in the magnetic susceptibility are just resolvable at slightly lower temperature. 
    
    Finally, in scenario $C$, in the range $-2\leq \mu \leq \mu_a$, a single first-order transition is observed. Included in Fig.~\ref{fig:scenarios} is an isopotential at $\mu=-1.4$, which shows a discontinuous jump in both observables without any apparent singularity in the response functions. Below $\mu=-2$, no further transitions are observed, %as expected.  
    the vacuum ($\eta_i = 0$) being the ground state.
    More detailed results for the three scenarios are given below.

    \begin{figure}
        \centering
         \includegraphics[scale = 1]{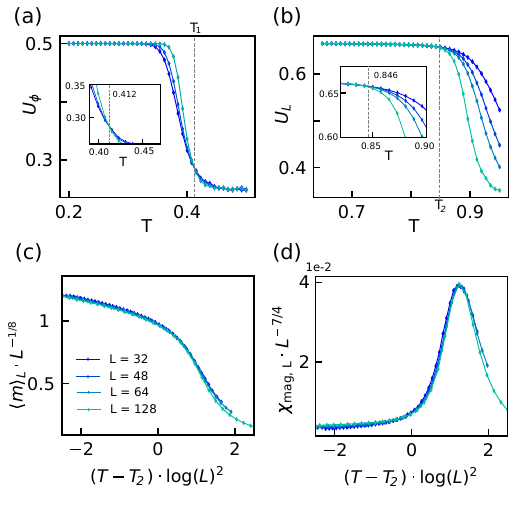}
        \caption{
        \textbf{Scenario A: two BKT transitions.
        Monte-Carlo simulations. } $\mu = 1 > \mu_t$.
        System size $L=32$ to $128$ (blue to green).
        (a) Low temperature BKT transition ($T_1\simeq 0.412$) obtained using the scale invariant point of the orientational Binder cumulant $U_\phi$, Eq.~\eqref{eq:Uphi}.
        (b) High-temperature BKT transition  ($T_2\simeq 0.846$) obtained using the scale invariance of the Binder cumulant $U_L$, Eq.~\eqref{eq:UL}.
        (c) BKT scaling behavior of $\langle m\rangle$ around $T_2$, %with the exponent of the BKT transition $2-y_h=1/8$ and the logarithmic scaling with system size, 
        see Eq.~\eqref{eq:scaling_m}. 
        (d) BKT scaling behavior of $\chi_\mathrm{mag}$ around $T_2$, see Eq.~\eqref{eq:scaling_chi}. 
        }
        \label{fig:lowNuMC}
    \end{figure}

    \begin{figure}
        \centering
        \includegraphics[scale = 1.]{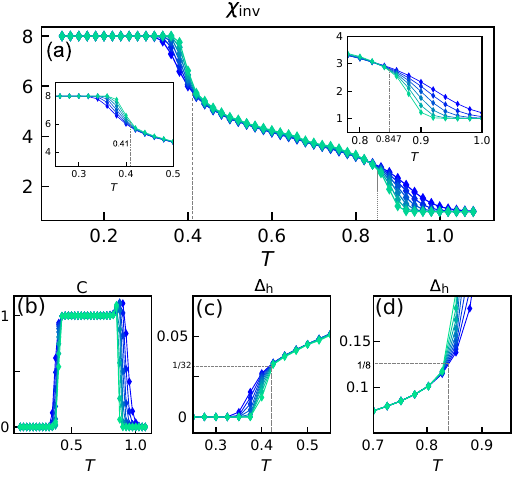}
        \caption{
        \textbf{Scenario A: two BKT transitions.
        Tensor networks (TNR).} %(NNR-TNR) at  .}
        $\mu = 1 > \mu_t$.
        Number of renormalization iterations from $7$ to $16$ (blue to green).
        (a) The gauge invariant $\chi$ {\it vs. } $T$, showing universal scaling between the two BKT transitions, giving $T_1\simeq 0.41$ and $T_2\simeq 0.847$. 
        (b) The central charge $c$ {\it vs.} $T$. Data is consistent with $c=1$ along the line of critical points, corresponding to free boson theory. 
        (c) The first scaling dimension $\Delta_h$ {\it vs. } $T$ near $T_1$. The universality with system size terminates at $T_1$ with $\Delta_h =1/32$.
        (d) $\Delta_h$ {\it vs. } $T$ with universal scaling up to $T_2$ where $\Delta_h = 1/8$.
        %corresponding to the known value for the BKT transition.
        %\todo[inline]{$\Delta_1\rightarrow \Delta_h$ above figs c snd d}
        }
        \label{fig:lowNuTNR}
    \end{figure}

    \subsection{High chemical potential: two continuous transitions} \label{ss:scenarioA}
    In the limit $\mu\to\infty$, our model reduces to the $p$-state clock model which exhibits two BKT transitions~\cite{Elitzur1979,Lapilli2006,Li2022},  as all the sites are filled ($\eta_i=1$) and the only degrees of freedom are the orientations $\theta_i$. The high temperature transition, at $T_2^{\infty} \simeq 0.89$, is  identical to that of the XY model~\cite{Lapilli2006}. The low-temperature transition is at $T_1^{\infty} \simeq 0.41$ as found from the duality argument~\cite{Jose1977,Savit1980,Elitzur1979,Ortiz2012,Lapilli2006,Li2022}. 
    %We show below that t
    This scenario also holds for finite $\mu$ down to $\mu_t$. %$\simeq -1.11$.
    %Here
    We focus on $\mu=1$ using both Monte-Carlo simulations and TNR.
    %, Fig.~\ref{fig:lowNuMC} and TNR, %Fig.~\ref{fig:lowNuTNR}.

    The high-temperature transition, $T_2$, is located from finite size scaling of the Binder cumulant $U_L$~\cite{Hasenbusch_2008}, Eq~\eqref{eq:UL}, 
    shown in Fig.~\ref{fig:lowNuMC}b.
    Within the QLRO phase, $U_L$ exhibits scale invariance and the cumulants at different sizes collapse over a finite temperature range.
    The transition point is associated with the disappearance of this scale invariance property. From Fig.~\ref{fig:lowNuMC}b, we obtain $T_2(\mu=1) \simeq 0.846$.
    The low-temperature transition is determined through the orientational Binder cumulant $U_\phi$~\cite{binder_cumulants_clock}, see Eq.~\eqref{eq:Uphi}.
    In the  QLRO phase, $m_\phi$ is continuously distributed on the unit circle, while in the LRO phase it is discretely distributed on the $p$ possible orientations of the clock model: $U_\phi$
    accounts for this shift from discrete to uniform distribution as it shows scale invariance
    at the low-temperature BKT transition $T_1$ only. Hence there is a conventional crossing point for the cumulants, as shown in Fig.~\ref{fig:lowNuMC}b.
    We estimate $T_1(\mu=1) \simeq 0.412$.

    Around the high-temperature transition, thermodynamics quantities have power law system size dependence with logarithmic corrections to scaling along the temperature axis~\cite{Kosterlitz1973}. The magnetic scaling exponent $y_h$ can therefore be extracted  from the finite-size scaling of the magnetization 
    and magnetic susceptibility,
      taking the logarithmic corrections into account~\cite{Bramwell1994},
    \begin{align}
        \langle m\rangle_L &\sim L^{2-y_h} \mathcal{M}\left([T-T_2(\mu)](\log L)^2\right), \label{eq:scaling_m} \\
        \chi_{\mathrm{mag}, L} &\sim L^{2(y_h - 1)} \mathcal{X}\left([T-T_2(\mu)](\log L)^2\right),\label{eq:scaling_chi}
    \end{align}
    with $y_h = 15/8$, where $\mathcal{M}$ and $\mathcal{X}$ are universal scaling functions. Monte-Carlo data, plotted in Fig~\ref{fig:lowNuMC}c-d, show excellent agreement with these scaling laws, confirming the BKT nature of the transition. 

    Using TNR,  we first compute the gauge invariant $\chi_\mathrm{inv}$, Eq.~\eqref{eq:chi_inv}.
    In Fig.~\ref{fig:lowNuTNR}a, the line of critical points between $T_1$ and $T_2$ is characterized by the scale invariance of $\chi_\mathrm{inv}$ with respect to the number of iterations of the algorithm~\cite{Li2022}. By estimating the termination the scale invariant region we obtain $T_1\simeq 0.41$ and $T_2\simeq 0.847$ in good agreement with the Monte-Carlo analysis.
    We also compute the CFT parameters.
    We plot the central charge $c$ in Fig.~\ref{fig:lowNuTNR}b, highlighting a line $c=1$ in the QLRO phase, consistently with CFT for free bosons~\cite{DiFrancesco1997}. As the phases below $T_1$ and above $T_2$ are not critical, we obtain $c=0$ in accordance with previous work on the $p$-state clock model~\cite{Li2022}.
    The magnetic scaling dimension $\Delta_h$ is displayed in Fig.~\ref{fig:lowNuTNR}c and d. Between $T_1$ and $T_2$, $\Delta_h$ varies with temperature and is independent of the number of iterations. The two extremes of scale-free behaviour, $T_1$ and $T_2$ are characterised by the predicted values for the BKT transitions, $\Delta_h=\frac{1}{32}$ and $\Delta_h = \frac{1}{8}$~\cite{Jose1977,Kosterlitz1973}.  This phenomenology, combining both numerical methods holds for all values $\mu>\mu_t$.

    \begin{figure}
    	\centering
    	\includegraphics[scale=1]{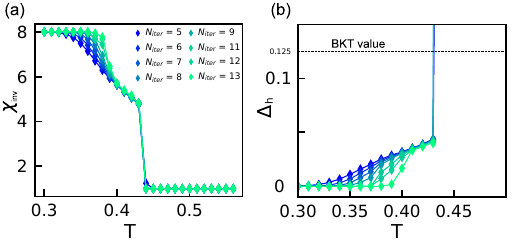}
    	\caption{  
    		\textbf{Scenario B: a BKT transition and a first-order transition.}
    		TNR for $\mu = -1.15 \in (\mu_t, \mu_a)$.
    		(a) Gauge invariant $\chi_\mathrm{inv}$ {\it vs.} $T$. Data shows universal scaling above the first transition at $T_1\simeq 0.4$ as in  Fig.~\ref{fig:lowNuTNR}a with a discontinuity at  $T_2=0.43$, characteristic of a first-order transition.
    		(b) $\Delta_h$ {\it vs.} $T$. Data shows universal scaling above $T_1$ as in Fig.~\ref{fig:lowNuTNR}c terminating at $T_2$ at a value significantly below
            $\Delta_h^{\mathrm{BKT}}=1/8$.
    	}
    	\label{fig:intermNu}
    \end{figure}

    \subsection{Intermediate chemical potential: one continuous and one discontinuous transition} \label{ss:scenarioB}
    %In-between the two previous scenarios, 
    Decreasing the chemical potential into the range $\mu_a < \mu < \mu_t$, the high temperature transition becomes discontinuous. 
    Using TNR, we first show that the low temperature transition remains of BKT type. % as in scenario A.
    This is highlighted by the continuity of the gauge invariant $\chi_\mathrm{inv}$ and its %logarithmic finite-size scaling
    dependence on system size which we anticipate to be logarithmic, see Fig.~\ref{fig:intermNu}a.
    Increasing the temperature leads to the second transition, \textcolor{blue}{at $T_2$,} where the gauge invariant is discontinuous. 

    The scaling dimension $\Delta_h$, Fig.~\ref{fig:intermNu}b, shows scale-invariant behavior between the two transitions $T_1<T<T_2$, as expected for the QLRO phase.
    Above the first-order transition $T_2$, $\Delta_h$ becomes ill-defined since the disordered phase in not critical. 
    The value of $\Delta_h$ at $T_2$ is inferior to the BKT value: $\Delta_h < \Delta_h^\mathrm{BKT} = 0.125$, highlighting the fact that a putative BKT transition is preempted by the first-order transition, as shown in the Hartree-Foch calculation.

    \begin{figure}
    	\centering
    	\includegraphics[scale = 1]{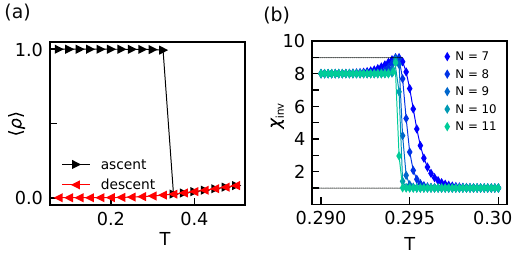}
    	\caption{\textbf{Scenario C: single first-order transition.}
    		$\mu=-1.4 \in (-2, \mu_a)$.
    		(a) $\langle \rho \rangle$ {\it vs.} $T$ from MC. Data show a large hysteresis loop  between configurations initialized in the disordered state with decreasing $T$ and configurations initialized in the ordered region with increasing $T$. 
    		(b) $\chi_{inv}$  {\it vs.} $T$ from TNR. Asymptote values are: $\chi_\mathrm{inv}=p=8$ in the broken symmetry phase, $\chi_\mathrm{inv}=1$ in the disordered phase and $\chi_\mathrm{inv}=p+1 = 9$ at the phase coexistence point. The transition is estimated at $T^\ast = 0.2943$. 
    		% (c) Scaling pour la 1ère ordre ??
    	}
    	\label{fig:highNu}
    \end{figure}

    \subsection{Low chemical potential: a single discontinuous transition} \label{ss:scenarioC}
    In scenario $C$,
    Monte-Carlo simulations show strong hysteresis through the transition, which is characteristic of metastable states, see Fig.~\ref{fig:highNu}a. When starting from a fully-occupied and ordered state at zero temperature and increasing the temperature, we observe a discontinuous jump in the density $\langle\rho\rangle$ to the low density disordered phase. Going in the other direction, the low density phase persists down to zero temperature.
    While these Monte-Carlo simulations strongly suggest a single discontinuous transition, they do not enable us to locate it precisely.
    
    Turning to the TNR method, we compute the gauge invariant $\chi_\mathrm{inv}$, Eq.~\eqref{eq:chi_inv}, which characterizes the number of competing phases. As shown in Fig.~\ref{fig:highNu}b,  it jumps from $\chi_\mathrm{inv}=1$ at high temperature, characteristic of the disordered phase, to $\chi_\mathrm{inv}=8$ corresponding to $\mathbb{Z}_8$ symmetry of the clock model. The transition itself is a singular point with $\chi_\mathrm{inv}=9$, corresponding to phase equilibrium between the $p=8$ ordered phases and the disordered phase. The discontinuity sharpens with the number of iterations of the TNR, allowing us to  precisely locate the transition point at $T^\ast = 0.294$ for $\mu = -1.4$. The  discontinuous transition is similar to that of the Blume-Capel model above the tricritical point~\cite{Blume1966,Capel1966,Blume1971,Kwak2015,Zierenberg2017,Moueddene2024} apart from the finite slope for the phase boundary.
    Extrapolating this result  to the zero temperature intercept at $\mu=-2$ gives a slope, $\frac{d\mu}{dT}=2.039$, in close agreement with the estimate from Eq.~\eqref{Clapeyron}. For $\mu < -2$, the system approaches a vacuum at low temperature and there is no further transition.

    Note that a detailed analysis of the first-order transition between QLRO and SRO phases as shown for example in Fig.~\ref{fig:intermNu}a, yields a similar step in $\chi_\mathrm{inv}$. In this case the evolution through the transition is from a universal value in the critical phase, $\tilde{\chi}$, to $\tilde{\chi}+c$ at the first-order transition to one in the disordered phase, where $c$ is a constant of order unity - see App.~\ref{app:tricritial_sm} and Fig.~\ref{fig:SM_tricritical}.

    \section{Phase diagram and tricritical point} \label{s:tricritical}

    \subsection{Phase diagram in the $(T, \mu, h=0)$ plane}
    
\begin{figure}
        \centering
            \includegraphics[width=\columnwidth]{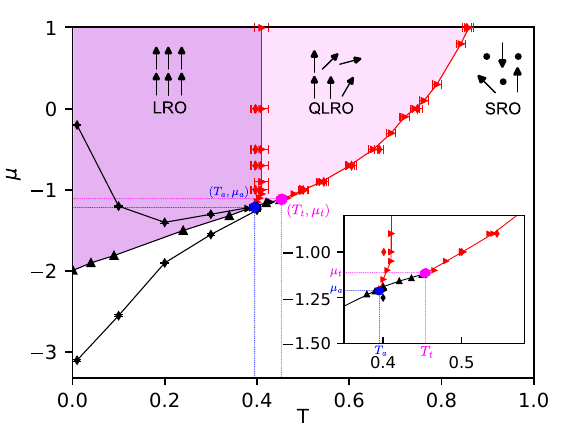}
        \caption{\textbf{Quantitative phase diagram.}
        The triangles correspond to transition points from TNR and diamonds from MC: red for a BKT transition and black for a first-order transition. Lines of appropriate color guide the eye. For MC %we show %we now have
        the two black lines correspond to the limits of hysteresis. The zones are colored: purple long-range order (LRO),  pink quasi-long-range order (QLRO) critical points, white  disorder (SRO). Also shown are
        the tricritical point (magenta) and the intersection of the low-$T$ transition with the first-order line (blue). 
        %{\bf Inset}  E
        Inset: Expanded region around the two special points.
        Again, the magenta area is a plane of critical points with QLRO order.}
        \label{fig:phaseDiag}
    \end{figure}

    Using the above methods, following both isopotentials and isotherms one can build the 
    phase diagram for zero field shown in Fig.~\ref{fig:phaseDiag}. This includes 
    the location of the transitions between LRO, QLRO and SRO phases, as well as their nature: BKT-type and first-order.
   
   The phase diagram includes two special points. %One corresponds to
   The first is the intersection of the low temperature BKT transition line with the first-order line, occurring at $(T_a, \mu_a)\simeq (0.397, -1.22)$. The line is almost vertical but shows a slight curvature near this point due to the variation in density as the first-order line is approached (see inset of Fig.~\ref{fig:phaseDiag}). This feature is also reproduced qualitatively in the Hartree-Foch phase diagram Fig.~\ref{fig:HartreeFock}.
   The second special point is the tricritical point~\cite{Cardy1979,Dillon2010,Santos2018,Skovdal2023} which we investigate below.

    \subsection{Tricritical point}
    \label{ss:tricritical}
While similar at first sight, the tricritical point here differs 
from that of the Blume-Capel model. The BKT transition is characterized by a single critical exponent whose root is the the scaling dimension $\Delta_h$ giving non-trivial eigenvalue exponent $y_h=d-\Delta_h$. This contrasts with a conventional second order transition that has two independent exponents associated with the field and the temperature axes $\Delta_h$ and $\Delta_t$ with $t$ the reduced temperature. Yet despite this difference, $\Delta_h$ takes the same value, $\frac{1}{8}$ for both BKT and two-dimensional Ising transitions~\cite{Berezinskii1971,Kosterlitz1973,Li2022}. The tricritical point introduces a further dimension to the renormalisation group along eigenvector $g=\mu-\mu_t + at$ with $t=T-T_t$ \cite{Kwak2015,Moueddene2024}, for which the eigenvalue and scaling dimension are $y_g^t=d-\Delta_g^t$. This makes three independent exponents for the Blume-Capel tricritical point and two for our model, since there is no thermal exponent. %two in the present case.
Although there is some literature on this subject~\cite{Vanderzande1988}, the values of these exponents remains an open question.

\begin{figure}
        \centering 
        \includegraphics[scale=1]{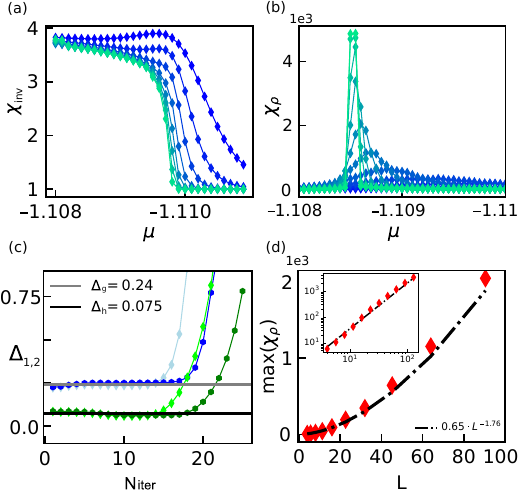}
        \caption{\textbf{BKT to first-order tricritical point.}
        (a)  $\chi_\mathrm{inv}$ {\it vs.} $\mu$ for $T=0.457$. The data shows a peak 
        in $\chi_\mathrm{inv}$ which scales away as the number of iterations increases, compatible with a tricritical point.
        (b) $\chi_\rho$ {\it vs.} $\mu$ for the same temperature. The data shows an emerging divergence which grows with increasing number of iterations for $\mu \simeq 1.1085$, consistently with 
        the estimate from the gauge invariant.
        (c) Scaling dimensions in the
        tricritical region as a function of TNR steps for
        $T=0.452, \nu=1.1185$ (lightgreen and lightblue) and $T=0.455, \nu = 1.1130$ (green and blue).  Values for the tricritical point the are estimated from the plateau regions.
        The first is
        %We observe a first tricritical value 
        $\Delta_h^t\simeq 0.075 \pm 0.01$, consistent with the Ising tricritical exponent. 
        The second is 
        $\Delta^t_2 \simeq 0.24 \pm 0.01$.
        (d) The maximum of the density susceptibility, $\chi^{max}_{\rho}$ as a function of $L=2^{N_{iter}}$ (red diamonds) compared with the scaling law $\chi^{max}_{\rho}=\chi_0 L^{d - \Delta_2^t}$ (dot-dashed line) -- see text.
       % Analysis of the divergence of density fluctuations as a function of the lattice linear size, compared to the power law $L^{d - \Delta_2^t}$.
        }
        \label{fig:tricritical}
    \end{figure}

The tricritical point corresponds to the point at which the peak of the gauge invariant from the TNR associated with a first-order transition vanishes, see  section~\ref{ss:scenarioC}, App.~\ref{app:tricritial_sm} and Fig.~\ref{fig:SM_tricritical}. From this we are able to make an accurate estimate its location at $T_t = 0.4545\pm 0.003$, $\mu_t = -1.1137 \pm 0.005$, as shown in Fig.~\ref{fig:tricritical}a.
A second estimate can be made from the density susceptibility $\chi_{\rho}$, see Eq.~\eqref{eq:chi_rho} which is finite along the BKT line but diverges non-trivially at this point, and scales as $\sim L^{d}$ along the first-order line. Although less precise, this divergence gives us an estimate
%our best estimate from this divergence gives
$T_t\approx 0.457$, $\mu_t\approx -1.114$, in good agreement with the first method, see Fig.~\ref{fig:tricritical}b. 

The scaling dimensions at the tricritical point can be extracted from TNR, see Fig.~\ref{fig:tricritical}c.
They approach stable asymptotes as the number of iterations increases before the renormalisation procedure pushes them away from the unstable fixed point. 
The smallest exponent, which we interpret as the magnetisation exponent is $\Delta_1 = \Delta_h^t = 0.075\pm 0.01$. This is consistent with the Blume-Capel tricritical exponent in two dimensions, $\Delta_h^{BC}=\frac{3}{40}$~\cite{DiFrancesco1997,Moueddene2024}. The second exponent which controls the density fluctuations is found to be $\Delta_2 = \Delta_{g}^t = 0.24\pm 0.01$ and is quite different from that of the Blume Capel model, $\Delta_g^\mathrm{BC} = 6/5 $~\cite{DiFrancesco1997,Kwak2015}. 

As a confirmation of this value for $\Delta_2$, we show in Fig.~\ref{fig:tricritical}d  a good agreement between the maximum of the susceptibility at the tricritical point and the scaling law $\chi_{\rho}^\mathrm{max}(L)\sim \chi_0L^{y_g^t}$, with $y_g^t=1.76$ corresponding to $\Delta_g^t=0.24$. The system size is extracted from the number of TNR iterations giving a large range of values. The scaling law that we propose is adapted from
finite size scaling close to the Blume-Capel tricritical point~\cite{Moueddene2024}. By setting the thermal scaling dimension equal to zero the thermal eigenvalue $y_t$ \cite{Moueddene2024} equals the spatial dimension.  Assuming that the amplitude of the bulk term is zero, it then follows that the leading contribution to the scaling of $\chi_\rho$ is the correction to scaling for Blume-Capel, $\chi_\rho^\mathrm{max}(L) \sim L^{y_{g}^t}$.

Our results are therefore compatible with the existence of a specific tricritical universality class for the BKT to first-order tricritical point.
%Further work will nevertherless be required to confirm this universality and establish accurate values for the tricritical exponents.
One of the exponents appears close to that for the Blume-Capel tricritical point, while the second, $\Delta_2\approx 0.24\approx\frac{1}{4}$ is different. More work is required both to confirm the scaling phenomenology used and to establish accurate values for the exponents.
%\todo[inline]{J'ai racourci, on peut supprimer.}

    \begin{figure*}
    	\centering
    	\includegraphics[width=\textwidth]{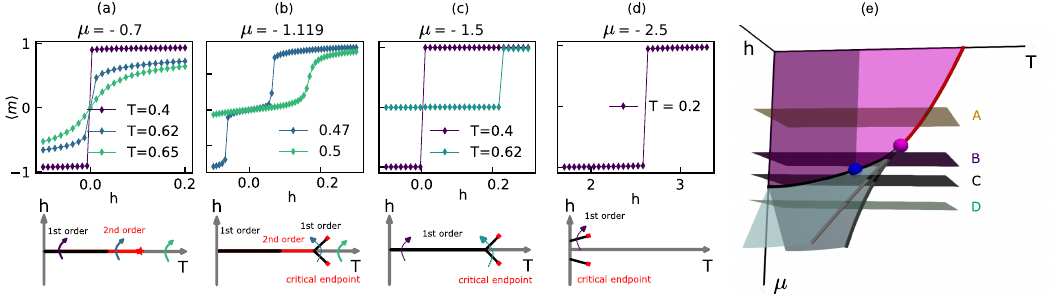}
    	\caption{
        \textbf{Cuts for non-zero field.}
        Top: magnetization from TNR. 
    		Bottom: phenomenological phase diagram. (a) $\mu > \mu_t$ (scenario A), (b) $\mu_t > \mu > \mu_a$ (scenario B), (c) $\mu_a >\mu > -2$ (scenario C), (d) $ -2 > \mu$, with first-order transition at non zero magnetic field. (e) Complete phase diagram in $(T, \mu, h)$ space with the different scenarios as cuts in this parameter space. The magenta area is a plane of critical points with QLRO order.
        }
    	\label{fig:plane_Th}
    \end{figure*}

    \subsection{Non-zero magnetic field}
    \label{ss:with_field}
    So far we have focused on the $h=0$ plane. Returning to the full Hamiltonian, Eq.~\eqref{eq:H}, and the complete expression of the initial tensor, Eq.~\eqref{eq:local_tensor}, we can map out the complete $(T, \mu, h)$ phase diagram.
Fig.~\ref{fig:plane_Th} shows temperature-field  cuts for four different values of the chemical potential $\mu$. The magnetization $\langle m\rangle$ is taken as the projection of the vector magnetization onto the field direction and is computed from TNR.

Cut (a) 
is in the KTHNY region with the LRO, QLRO and SRO phases lying along the zero field axis. The QLRO phase constitutes a continuous line of gaussian phase transitions with exponent $\frac{1}{32}<\Delta_h(T)<\frac{1}{8}$. The upper panel shows $\langle m\rangle$ against field for isotherms in each region. There is a discontinuity  on crossing the first-order line at low temperature, a critical response crossing the gaussian line and a paramagnetic response at high temperature. 
Near the gaussian critical point we expect $\langle m\rangle\sim h^{1/\delta(T)}$, with $\delta(T)=\left(\frac{d-\Delta_h(T)}{\Delta_h(T)}\right)$ with $15<\delta(T)<63$, leading to an extremely rapid evolution of $\langle m\rangle$ for small field, as confirmed by our data. 

Cut (b)
is in the intermediate regime with, in zero field, a first-order transition into the QLRO phase. An isotherm crossing the the line of gaussian transitions should show a critical response, as for cut (a). However, the large value of $\delta(T)$ puts the continuous evolution outside the resolution of our TNR procedure, making it appear as a discontinuity (not shown).  The lowest temperature isotherm shown is above the zero field transition temperature. It encounters a first-order jump in the magnetization at finite $h$ and is consistent with the prolongation of the first-order transition into the broken symmetry phase. 
The first-order wings terminate in  liquid-gas like critical end points as illustrated in the figure. The higher temperature isotherm is above this temperature and a continuous response to the field is indeed observed. 

As the tricritical point is approached the wings %will
close up. The two critical end points %will be
are superposed with the zero field transition which becomes a BKT transition. The data therefore illustrates that, apart from the particularities of the BKT transition, this point has the same thermodynamic construction as a regular tricritical point. 

Cut (c)
is in the regime with a single first-order transition at zero field, as shown by the low temperature isotherm in the lower panel. The second isotherm lies above the zero field transition temperature and crosses the two symmetry sustaining first-order lines which again terminate in liquid-gas like critical end points. 
Finally cut (d) 
lies below $\mu=-2$ so that there is no phase transition at zero field. However, isotherms at sufficiently low temperatures encounter two first-order lines extrapolating to finite field at zero temperature, as illustrated in the lower panel. The ensemble of results are illustrated in Fig.~\ref{fig:plane_Th}e showing a winged phase diagram characteristic of tricriticality~\cite{Plascak1993,Taufour2016,Raban2019} while also including the QLRO phase in the zero field plane.

    %%%%%%
    \section{Discussion} \label{s:discussion}

    \subsection{Melting scenarios for realistic systems}

\begin{figure}
        \centering 
        \includegraphics[width=\columnwidth]{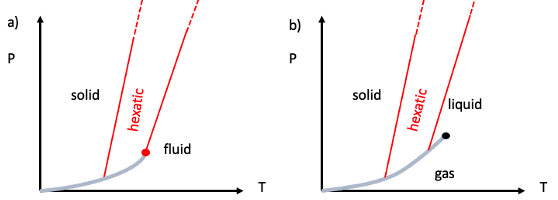}
        \caption{\textbf{Pressure-temperature phase diagrams for 2D particles.} (a) Without attractive interactions, we expect a phase diagram similar to that for our model with the vapor pressure $P$ replacing the chemical potential $\mu$. (b) With attractive interactions, an additional liquid-gas transition appears, it could be reproduced by a BEG term in our Hamiltonian.}
        \label{fig:P-T}
    \end{figure}

    For a 2D fluid in contact with a bulk ideal gas, 
    %as for Langmuir films,
    the intensive variable can be interchanged between the chemical potential $\mu$ and the vapour pressure $P$: $P=P_0\exp{(\beta\mu)}$, where $P_0$ is a reference pressure from which the (excess) chemical potential is measured~\cite{Colella1986}.
    We therefore expect the $(\mu, T)$ phase diagram of the lattice model in zero field, Fig.~\ref{fig:phaseDiag}, to be a good indicator of the behavior of realistic systems in the $(P, T)$ plane. As a consequence we are able to sketch a possible phase diagram for 2D spheres in Fig.~\ref{fig:P-T}a. %hold qualitatively for continuous systems. 
    It exhibits the three phases: solid, hexatic and fluid, corresponding to LRO, QLRO and SRO. It  also contains the tricritical point separating BKT and first-order transitions from the disordered fluid phase and the dense, QLRO ordered phase as well as the intercept of the first-order line with the low temperature BKT transition. The figure highlights the fact that, apart from special cases the melting scenarios along  particular isobars or isotherms are not unique~\cite{Nelson1979,Hajibabaei2019,Toledano2021,Tsiok2022}.
    
 Conspicuous by its absence in this figure is a liquid-gas phase transition. This is rectified in Fig.~\ref{fig:P-T}b, where the tricritical point is replaced by a second intersection between BKT and first-order lines. Beyond this point the line divides the fluid phase into high density liquid and low density gas phases. The line ends at a critical end point. The existence of a liquid-gas transition depends on the presence of an attractive interaction and is non-universal~\cite{BarratBook2003}. Such a transition could be generated in the lattice model by including a bi-quadratic coupling
\begin{equation}
    -K\sum_{<i,j>} \eta_i^2\eta_j^2,
\end{equation}
    with $K>0$ defining a generalised Blume-Emery-Griffiths model \cite{Blume1971}. Early theoretical work on vector models \cite{Cardy1979} suggested that the threshold for evolution from a tricritical point to liquid-gas coexistence would be $K_c=0$ or very close to it. However, more recent numerical work suggests a threshold of finite strength \cite{Santos2018}.

    Although Figs.~\ref{fig:P-T}a and b may capture the behavior of many systems they are not exhaustive.
    For example, %adding a short range attractive potential to a hard core repulsion leads to melting scenario
    assemblies with short-range attractions melt via two first-order transitions \cite{Bladon1995}.
    Such a scenario has been reported in colloidal systems \cite{Marcus1996}.
    The lower temperature transition is reported to terminate in a solid-solid critical end point opening the door to more complex melting scenarios at higher pressure. 

A few words are in order concerning
scale-free potentials, including hard discs, which offer a unique melting scenario \cite{Kapfer_2015}. In this case, there is a single intensive thermodynamic variable, $x={P}/{T}$ such that the phase boundaries are straight lines which intercept at the origin. %, $P=0,T=0$.
However, %approximating this with 
for a realistic potential, one would ultimately expect a scale invariant approximation to break down, giving a crossover to a richer phase diagram as in Fig.~\ref{fig:P-T}a or b.

   \subsection{Experimental systems}
      
The biggest asset of our on-lattice approach to two-dimensional melting is that it gives access to the full phase diagram, including application of a field conjugate to the orientational order parameter. Such completeness remains a monumental task for both experimental and numerical approaches. 
Candidate experimental systems include free-standing liquid crystal films, electrons absorbed into helium films, noble gases adsorbed onto graphene and other surfaces and sub-micron sized colloids trapped between glass plates of varying thickness (see Ref.~\cite{Strandburg1988} and references therein). In all cases complexity and measurement constraints limit the extent of comparison with model systems. Mapping results onto our phase diagram should help in gauging this comparison. 

In free standing bi-layer liquid crystal films an intermediate hexatic phase with quasi-long range order \cite{Cheng1988} has been observed which suggest that the passage from smectic-$A$ to $A^{\ast}$ to smectic-$B$ could follow the KTHNY scenario. However, film stability and competing phases add to the experimental complexity. The electron systems have long range Coulomb interactions and so could show KTHNY melting but  orientational ordering is difficult to observe \cite{Strandburg1988}. Of the adsorbed systems xenon on graphene is of particular interest. Experiments performed at fixed density show that for sub monolayer coverage melting is strongly first-order, becoming continuous as the coverage increases. Vapor pressure isotherms have been measured in this region \cite{Colella1986}, showing that the evolution could possibly be via a tricritical point in close analogy with our model. Experiments on colloidal systems \cite{Murray1987,Marcus1996,Han2008} show evidence of both two stage melting \cite{Murray1987} and first-order \cite{Marcus1996} scenarios depending on density and interaction potential and some temperature control is possible \cite{Han2008} but the analysis of an extensive phase diagram would be extremely challenging.

\subsection{Model systems off and on lattice}

For practical reasons, numerical simulations on model fluids have been performed almost exclusively in the canonical ensemble with fixed volume and particle number \cite{Strandburg1988,Kapfer_2015}. This is particularly true at high density where particle insertion or particle rescaling methods become technically challenging \cite{FrenkelBook2023}. 
Working in the canonical ensemble rather than in an ensemble with a freely fluctuating order parameter makes analysis of phase transitions harder. As a consequence, as we have shown  the on-lattice procedure that we propose opens the door to a more complete and detailed analysis of these properties. 

In particular, the simplicity of the model and the recent advances in tensor network renormalization group have allowed for a detailed study of the tricritical point connecting BKT and first-order transitions. 
It would be almost impossible to approach it without the the use of TNR on the lattice model. Even here, more work is required to fully characterize this point. We have made a scaling hypothesis that the tricritical properties emerge from finite size scaling of the Blume-Capel tricritical point, taking the limit that the temperature scaling dimension goes to zero. As a consequence the corrections to scaling become the leading effects of the BKT tricritical point. Our numerical results are consistent with a field scaling dimension, $\Delta_h^t$ which is the same as at the Blume-Capel point and a second exponent, $\Delta_g^t$, related to the chemical potential which is quite different. More work is required to test this hypothesis further, for example for vector models and to verify the results for clock models for different values of $p$.

Finally, there are limits to the analogy between the lattice model and two-dimensional fluid systems. At first sight the topological objects that deconfine at the two BKT transitions appear different in the two systems. In the lattice model, at the upper transition vortices in the spin field with emergent continuous symmetry deconfine, as in the $XY$ model \cite{Kosterlitz1973,Ortiz2012}. At the lower transition the BKT transition is identified through a duality transformation \cite{Jose1977,Ortiz2012} between temperature and inverse temperature and between fugacities for vortices and for the relevance of the $p$-fold perturbation. Topological defects are therefore in the thermodynamic variables of the dual model, suggesting that the low temperature phase is a deconfined phase analogous to the high temperature, disordered phase. This seems at odds with the melting scenario in which dislocations and disclinations progressively deconfine on heating. As translational symmetry is integrated out of the problem in the lattice model, there is no obvious analogue of the Burgers vectors characterising the deconfined dislocations of the hexatic phase, although they may exist in the emergent $U(1)$ field that drives the transition. In future work it would be interesting to pursue these questions through a detailed study of the dual model. 
%\fi

    %%%%%%
    \section{Conclusion} \label{s:conclusion}
    Motivated by the study of two-dimensional melting, we introduced a minimal on-lattice model: a clock model with vacancies: spins have $p=8$ possible orientations and two allowed lengths, $0$ or $1$. The model has three control parameters: the temperature $T$, the chemical potential $\mu$ and external field, $h$ coupling to the orientational order parameter. We have accessed the complete phase diagram using a combination of Monte Carlo simulation and tensor network renormalization group, with the main features confirmed via variational mean field theory and thermodynamic arguments. 

    The model captures an extensive set of possible scenarios for two-dimensional melting and highlights the non-universal nature of this process for most realistic systems. For large $\mu$, the model 
reduces to the clock model~\cite{Elitzur1979,Lapilli2006,Li2022} and exhibits two BKT transitions separating three phases with long-range order, quasi-long-range order  and short-range order in analogy with the KTHNY scenario of melting~\cite{Halperin1978,Nelson1979,Young1979,Kapfer_2015,Prestipino2011,Durand2019}.
For intermediate $\mu$ the 
high-temperature transition becomes discontinuous, as is the case for hard spheres ~\cite{Bernard_2011,Engel2013,Qi2014,Kapfer_2015}. Finally, for low $\mu$, the system disorders via a single discontinuous transition as can occur in more complex models ~\cite{Saito1982,Chui1983,Prestipino2012,Dudalov2014}.

The evolution of the high temperature transition from BKT to first-order is via a special tricritical point \cite{Cardy1979,Vanderzande1988,Dillon2010} which we have accessed in detail using tensor network renormalisation group. It is possible that this point is of experimental relevance for two-dimensional melting \cite{Colella1986}, films of $^3$He-$^4$He mixtures and magnetic metamaterials~\cite{Skovdal2023} and will be present in many model fluid systems. 

Making small changes to the model, a liquid gas phase phase equilibrium line could emerge from the tricritical point, terminating in a critical end point and giving further possibilities to the phase diagram. We therefore hope that the lattice gas picture helps provide a global vision of two dimensional melting and will lead to more comprehensive numerical studies and give directions for future experiments.  

\begin{acknowledgements}

We thank Steven Bramwell, Michael Faulkner, Andrei Fedorenko and Pascal Viot for useful discussions. PCWH and AR thank the French American Center for Theoretical Science, CNRS, KITP, University of California, Santa Barbara, for financial support. The work was supported 
in part by NSF Grant No.
PHY-2309135 to the Kavli Institute for Theoretical Physics
(KITP).

\end{acknowledgements}

    \appendix

    \section{Variational principle}
    \label{app:Hartree}
    
    \subsection{Derivation of self-consistent equations}

    From the approximate Hamiltonian, Eq~\eqref{eq:approx_mf}, we can compute the associated partition function $\mathcal{Z}_\mathrm{MF}$ and free energy $\Omega_\mathrm{MF}=-\frac{1}{\beta}\log \mathcal{Z}_\mathrm{MF}$. We obtain
    \begin{align}
        %\mathcal{Z}_\mathrm{MF} &= e^{-\beta \mathcal{H}_0}(1+ e^{\beta \mu_\mathrm{eff}})^N \sum_{\{ \theta_i\}}e^{-\frac{\beta J_\mathrm{eff}}{2} \sum_{\langle i,j \rangle} (\theta_i - \theta_j)^2}, \\
        \Omega_\mathrm{MF} &= \mathcal{H}_0 +  \Omega_\mathrm{gauss} - \frac{N}{\beta}\log(1 + e^{\beta \mu_\mathrm{eff}}), \\
        \Omega_\mathrm{gauss} &= -\frac{1}{\beta} \log\left(\sum_{\{ \theta_i\}}e^{-\frac{\beta J_\mathrm{eff}}{2} \sum_{\langle i,j \rangle} (\theta_i - \theta_j)^2})\right).
    \end{align}
    %\todo[inline]{ $F_\mathrm{gaussian}$ not defined}
    
    We use the variational principle for the exact free energy $\Omega$~\cite{Feynman2018},
    \begin{equation}
        \Omega \leq \Omega_\mathrm{MF} + \langle \mathcal H - \mathcal{H}_\mathrm{MF}\rangle_\mathrm{MF},
        \label{eq:hartree1}
    \end{equation}
    where $\langle\rangle_\mathrm{MF}$ denotes an average with respect to the canonical law associated with $\mathcal{H}_\mathrm{MF}$.
\iffalse
    We obtain
    %\begin{equation}
    %\begin{aligned}
    \begin{multline}
    \Omega
    \;\le\; %&
    \Omega_{\mathrm{MF}}
    - J \sum_{\langle i,j\rangle}
      \big\langle \eta_i \eta_j \cos(\theta_i - \theta_j) \big\rangle_{_\mathrm{MF}}
    \\ %[4pt]
    %&\quad
    - \mu \sum_i \big\langle \eta_i \big\rangle_{\mathrm{MF}}
    - \mathcal{H}_0
    %\\ %[4pt]
    %&\quad
    - \frac{J_{\mathrm{eff}}}{2}
    \sum_{\langle i,j \rangle}
    \big\langle (\theta_i - \theta_j)^2 \big\rangle_{_\mathrm{MF}} \\
    + \mu_{\mathrm{eff}} \sum_i \big\langle \eta_i \big\rangle_{\mathrm{MF}} .
    \end{multline}
    %\end{aligned}
    %\end{equation}
\fi
    Let the right hand side of Eq.~\eqref{eq:hartree1} be $\mathcal{G}$. %such that
    Since the variables $\{\eta_i, \theta_i\}$ are decoupled %variables 
    in the variational Hamiltonian, we obtain
    %Eq.~\eqref{eq:approx_mf}, we obtain
    %\begin{equation}
    %\begin{aligned}
    \begin{multline}
    %\Omega
    %\;\le\; %&
    \mathcal{G} = 
    \Omega_{\mathrm{MF}} - \mathcal{H}_0
    - J \,\langle \rho \rangle_\mathrm{MF}^{2}
    \sum_{\langle i, j\rangle}
    \big\langle \cos(\theta_i - \theta_j) \big\rangle_\mathrm{MF}
    \\ %[4pt]
    %&\quad
    - \frac{J_{\mathrm{eff}}}{2}
    \sum_{\langle i, j\rangle}
    \big\langle (\theta_i - \theta_j)^2 \big\rangle_\mathrm{MF}
    %\\ %[4pt]
    %&\quad
    - N(\mu - \mu_{\mathrm{eff}})\langle \rho \rangle_\mathrm{MF}, %\,
    \end{multline}
    %\end{aligned}
    %\end{equation}
    where $\langle \rho \rangle_\mathrm{MF} = \langle \eta_1 \rangle_\mathrm{MF} $ is the average density.
    %Let us call the second member $\mathcal{G}$ such that $\Omega \leq \mathcal{G}.$
    Requiring $ \frac{\partial \mathcal{G}}{\partial J_\mathrm{eff}} = 0$ leads to
  
    \begin{equation}
    \frac{\partial \Omega_\mathrm{MF}}{\partial J_\mathrm{eff}} 
        -\frac{Nz}{2}\left( J\langle \rho \rangle_\mathrm{MF}^2\frac{\partial C}{\partial J_\mathrm{eff} } + \frac{\sigma^2 }{2}
    +\frac{J_\mathrm{eff}}{2}\frac{\partial \sigma^2}{\partial J_\mathrm{eff}} \right)=0,
        \label{eq:minimization}
    \end{equation}
    where $z=4$ is the coordination of the lattice. $\sigma$ and $C$ are defined over a single bond $(i,i+\hat x)$ where $i+\hat x$ is the site on the right of i,
    \begin{align}
        %C &= \sum_{\langle i,j\rangle} \langle \cos(\theta_i - \theta_j)\rangle_\mathrm{MF}, &
    %\sigma^2 &= \sum_{\langle i,j\rangle} \langle (\theta_i - \theta_j)^2\rangle_\mathrm{MF}. &
    \label{eq:def_sigma2}
    \sigma^2 &=  \langle (\theta_i - \theta_{i+\hat x})^2\rangle_\mathrm{MF}, \\
        C &= \langle \cos(\theta_i - \theta_{i+\hat x})\rangle_\mathrm{MF} = e^{-\frac{\sigma^2}{2}}.
    \end{align}
    In the last line we used the fact that in the gaussian model, $(\theta_1-\theta_2)$ is normally distributed with variance $\sigma^2$.
   
    Also using gaussian statistics, we find
    \begin{equation}
        \frac{\partial \Omega_\mathrm{MF}}{\partial J_\mathrm{eff}} = \left \langle \frac{\partial\mathcal{H}_\mathrm{MF}}{\partial J_\mathrm{eff}} \right \rangle_\mathrm{MF} = \frac{1}{2}\frac{Nz}{2}\sigma^2,
    \end{equation}
    Eq.~\eqref{eq:minimization} simplifies into
    \begin{equation}
        J_\mathrm{eff} = J \langle \rho \rangle_\mathrm{MF}^2e^{-\sigma^2/2}.
        \label{eq:res_Jeff}
    \end{equation}
    %\todo[inline]{Vérifier (l'éq avait disparu)}

    In the same manner, requiring $ \frac{\partial \mathcal{G}}{\partial \mu_\mathrm{eff}} = 0$ gives
    \begin{equation}
        \mu_\mathrm{eff} = \mu + 2J\langle \rho \rangle_\mathrm{MF} e^{-\sigma^2/2}.
        \label{eq:res_mueff}
    \end{equation} 

    %\todo[inline]{Fin à revoir, pas clair.}
    The goal is now to compute $\sigma^2$, to close the system of equations. 
    We consider the angular part of the variational Hamiltonian
    $\mathcal{H}_\mathrm{gauss} = \frac{J_\mathrm{eff}}{2}\sum_{\langle i, j\rangle} (\theta_i-\theta_j)^2$.
    We define the Fourier space variables $\theta_{\vec q}$ as
    %Considering again $\mathcal{H}_0$, and going to Fourier space using 
    $\theta_j=N^{-1/2}\sum_{\vec q}\theta_{\vec q} e^{i\vec q\cdot\vec r_j}$ where $\rr_j$ is the position of spin $j$. The Hamiltonian is diagonal,
    %($\theta_{-q}=\theta_q^\ast$) we get:
    \begin{align}
      \mathcal{H}_\mathrm{gauss} &=\frac{J_\mathrm{eff}}{2} \sum_{\vec q} w_{\vec q}|\theta_{\vec q}|^2, \\
      w_{\vec q}&=2\,(2-\cos q_x-\cos q_y).
    \end{align}
    The gaussian propagator
    %, i.e. the correlation between the modes under the canonical law associated with $\mathcal{H}_\mathrm{gauss}$
    is
    $\langle\theta_{\vec q}\theta_{\vec q'}\rangle_\mathrm{gauss}
    =\delta_{\vec q',-\vec q}\, T/(J_\mathrm{eff} w_{\vec q})$ for $\vec q\neq0$. 
  
    We can now compute $\sigma^2$ from Eq.~\eqref{eq:def_sigma2},
    \begin{equation}
     %\sigma^2\equiv\langle(\theta_i-\theta_{i+\hat x})^2\rangle_0
      %=\frac{T}{J_\mathrm{eff}}\int_{[-\pi,\pi]^2}\!\frac{d^2q}{(2\pi)^2}\,
      %\frac{1-\cos q_x}{\,2-\cos q_x-\cos q_y\,}
      \sigma^2 = \frac{T}{J_\mathrm{eff}}\int_{[-\pi,\pi]^2}\frac{d^2\vec q}{(2\pi)^2} \frac{2(1-\cos q_x)}{w_{\vec q}} 
      = \frac{T}{2J_\mathrm{eff}},
      \label{eq:res_sigma2}
    \end{equation}
    where we have used the symmetry $q_x \leftrightarrow q_y$.

    We retrieve the result of the Hartree approximation for the XY model~\cite{Portelli2001}. Together, Eq.~\eqref{eq:res_Jeff}, \eqref{eq:res_mueff} and \eqref{eq:res_sigma2} lead to the self-consistent system, Eq.~\eqref{eq:system_eq}.

    \subsection{Supplementary curves for $\eta(\mu, T)$ and $\rho(\mu, T)$}
\begin{figure} %[h]
        \centering
    \includegraphics[width=\linewidth]{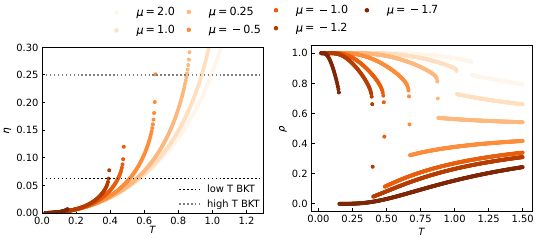}
        \caption{\textbf{Solutions of the self-consistent equations~\eqref{eq:system_eq}.} Left: anomalous exponent when varying the chemical potential. Right: density curves varying $\mu$. Note that the density curves always display a discontinuity.}
        \label{fig:SM_hartree}
    \end{figure}

    %\todo[inline]{Some text here to say look at the figure.}
    Starting from Eq.~\eqref{eq:system_eq}, we can numerically integrate the system to get $J_\mathrm{eff}, \nu_\mathrm{eff}$, and a fortiori $\langle \rho \rangle(T, \mu)$ and $\eta (T, \mu)$. We provide in Fig.~\ref{fig:SM_hartree} some results for these two observables for different values of $\mu$. This highlight the discontinuity in density, and the increasing sharpness in $\eta$ when approaching the first-order regime.

    \section{Tensor-network approach} \label{app:tnr}

    \subsection{Impurity method for local observables}
    \label{sec:impurity}
    
    Local observables are computed within the same tensor-network framework
    used for the partition function, by replacing a single local tensor with an
    \emph{impurity} tensor that carries the operator
    weight~\cite{MORITA201965}. We start from a local one-body observable: the thermal average of the magnetization in the $x$ direction -- the same method works for the average density $\langle\rho\rangle$. Using translational invariance,
    \begin{equation}
        \langle m \rangle = \langle S_0^x \rangle
        = \frac{1}{\mathcal{Z}}
          \sum_{\{\eta_i, \theta_i\}} S_0^x e^{-\beta\mathcal{H}(\{\eta_i, \theta_i\})}.
        \label{eq:s0_def}
    \end{equation}

    In the tensor-network representation the Boltzmann weight is encoded in a
    local tensor $T_{urdl}$, whose legs $(u,r,d,l)$ carry the bond indices to
    the up, right, down and left neighbours, and the partition function is the
    full network contraction $\mathcal{Z} = \mathrm{tTr}\bigl(\bigotimes_i
    T\bigr)$. The numerator of Eq.~\eqref{eq:s0_def} is obtained by inserting
    the factor $S_0^x$ into the local weight at the reference site, which defines
    a single impurity tensor $T^{\mathrm{imp}}_{urdl}$ sitting in an otherwise
    uniform network, leading to
    \begin{equation}
        \langle s_0 \rangle
        = \frac{\displaystyle
            \mathrm{tTr}\Bigl( T^{\mathrm{imp}} \otimes
            \bigotimes_{i \neq 0} T \Bigr)}
          {\displaystyle
            \mathrm{tTr}\Bigl( \bigotimes_{i} T \Bigr)} .
        \label{eq:impurity_ratio}
    \end{equation}
    The single impurity tensor is carried along through the coarse-graining
    flow exactly like the bulk tensors, so that
    Eq.~\eqref{eq:impurity_ratio} is evaluated with the same renormalized
    tensors used for $\mathcal{Z}$, at a cost comparable to the partition
    function itself.
    
    A practical caveat applies to magnetic observables. The contraction
    \eqref{eq:impurity_ratio} evaluates the \emph{exact} thermal average over
    the full, symmetry-preserving partition function. In the absence of a
    symmetry-breaking field ($h=0$), we therefore obtain a vanishing magnetization despite the spontaneous symmetry breaking below the critical temperature.
  
    For this reason we report direct magnetic observables
    only at finite field, see Sec.~\ref{ss:with_field}.
    
\begin{figure} %[h]
        \centering
        \includegraphics[width=0.7\linewidth]{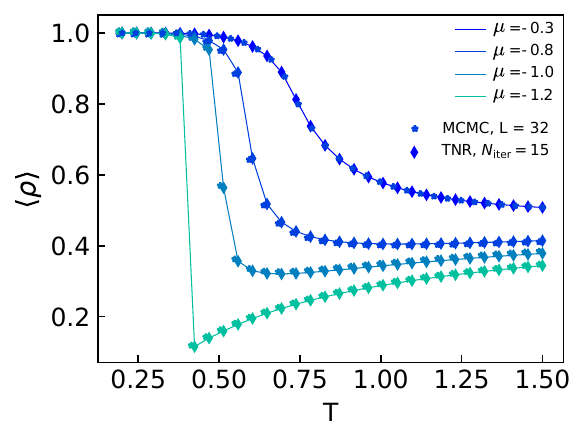}
        \caption{\textbf{Density from TNR.} Density evolution when decreasing $\mu$ with TNR compared with MCMC simulations, at fixed system size (see Fig.~\ref{fig:scenarios}). The results display a  behavior consistent with the  scenarii presented in the main text.}
        \label{fig:density_tnr}
    \end{figure}

    There is no such difficulty for the density $\langle\rho\rangle$. In Fig.~\ref{fig:density_tnr},
    %Finally, 
    we provide density profiles from the impurity method described above,
    in a similar fashion to Fig.~\ref{fig:scenarios}b. These results are in excellent agreement with Monte-Carlo simulations.

    Finally, given the field-dependent magnetization $m(T,\mu,h)$, the magnetic
    susceptibility follows from numerical differentiation,
    \begin{equation}
        \chi_{\mathrm{mag}}(T,\mu,h)
        = \left( \frac{\partial m}{\partial h} \right)_{\!T,\mu},
    \end{equation}
    and, in the same way, the density response (isothermal compressibility) is
    \begin{equation}
        \chi_{\rho}(T,\mu, h)
        = \left( \frac{\partial \rho}{\partial \mu} \right)_{\!T},
    \end{equation}
    where $\mu$ is the chemical potential controlling the density. By the
    fluctuation--dissipation relation these response functions equal the
    corresponding connected variances, providing an independent cross-check
    against Monte Carlo estimates.
    
    \subsection{CFT data}
    \label{sec:cft_data}

        \begin{figure*}
        \centering
        \includegraphics[width=0.8\textwidth]{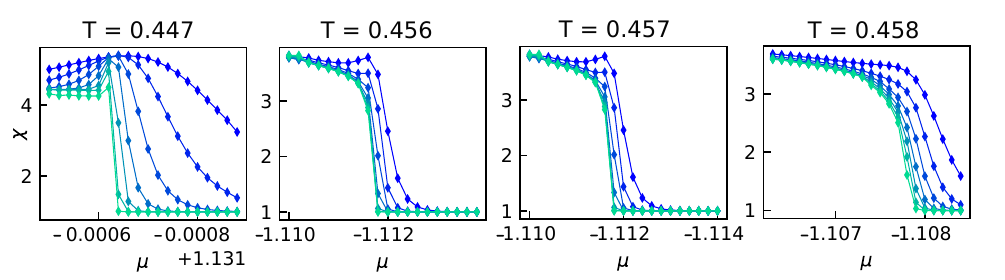}
        \caption{\textbf{Evolution of the gauge invariant peak around tricriticality.} $N_\mathrm{iter}$ from 7 (blue) to 15 (green). T = 0.447: still first-order. T = 0.458 : disparition of the peak, highlighting the disparition of the first-order transition.}
        \label{fig:SM_tricritical}
    \end{figure*}
    
    A central strength of TNR-type algorithms is their ability to extract
    conformal field theory (CFT) data---scaling dimensions and the central
    charge---directly from the coarse-grained tensors. We recall here only the
    main idea, developed in Refs.~\cite{Gu2009,evenbly2015}.
    
    Consider a two-dimensional classical system at criticality, described by a 2D CFT placed on a torus of typical length $L$. 
    One can show~\cite{Gu2009} that the partition function can be diagonalized as 
    \begin{equation}
        \mathcal{Z}
        %= \operatorname{tTr}\Bigl\{ \textstyle\bigotimes_i^\infty T_i \Bigr\}
        = \sum_\alpha
          e^{-\frac{2\pi\beta}{L}\left(\Delta_\alpha - \frac{c}{12}\right)},
        \label{eq:Z_spectral}
    \end{equation}
    where $\alpha$ runs over all primary fields \emph{and} their descendants, $\Delta_\alpha$ are the scaling dimensions and $c$ is the central charge. The term $-c/12$ is the universal Casimir contribution induced by the finite circumference~\cite{DiFrancesco1997}.

    The Gu--Wen prescription~\cite{Gu2009} makes the spectral content of
    Eq.~\eqref{eq:Z_spectral} directly accessible from a single coarse-grained
    tensor. Provided that the partition function reads $\mathcal{Z} \simeq \mathrm{tTr}(T^{\infty}$) where $T^{\infty}$ is the fixed-point tensor from the TNR procedure,
,
    one builds the transfer matrix
    \begin{equation}
        M_{ik} = \sum_{j} T^{\infty}_{ijkj},
        \label{eq:gu_wen_matrix}
    \end{equation}
    obtained by tracing one pair of opposite legs of $T^{\infty}$ periodically,
    so that $M$ propagates the system along the orthogonal direction over a
    single tensor width. 
  
    Using the scale invariance of $T^{\infty}$ along with Eq.~~\eqref{eq:Z_spectral}, one shows~\cite{Gu2009} that
    the eigenvalues of $M$ are %then
    \begin{equation}
        \lambda_\alpha
        = e^{-2\pi\left(\Delta_\alpha - \frac{c}{12}
                        + \mathcal{O}(1/L)\right)},
        \label{eq:gu_wen_eigenvalues}
    \end{equation}
    the $\mathcal{O}(1/L)$ corrections reflect the residual breaking of
    scale invariance at finite bond dimension. In practice the universal data
    are read off from \emph{ratios} relative to the leading eigenvalue
    $\lambda_0$, % (the identity, $\Delta_0 = 0$),
    which cancel the non-universal
    normalization,
    \begin{equation}
        \Delta_\alpha
        = -\frac{1}{2\pi}\,
          \ln\!\frac{\lambda_\alpha}{\lambda_0},
        \qquad
        \ln \lambda_0 = \frac{\pi c}{6} + \mathrm{(non-universal)},
        \label{eq:extract}
    \end{equation}
    so that the scaling dimensions follow from the eigenvalue ratios while the
    central charge is fixed by the leading eigenvalue together with the free
    energy normalization.
    
    Finally, as discussed in the main text, the standard RG eigenvalues $y_\alpha$
    %(relevant exponents) 
    are recovered from the scaling dimensions as
    \begin{equation}
        y_\alpha = d - \Delta_\alpha,
    \end{equation}
    with $d=2$ the spatial dimensionality. The relevant operators
    correspond to $\Delta_\alpha < d$, i.e. $y_\alpha > 0$.

    %\section{numerics: supplementary results}
    \subsection{A gauge-invariant indicator of spontaneous symmetry breaking}
    \label{sec:chi_inv}
    
    Upon convergence of the %TRG/
    TNR flow we obtain a fixed-point tensor
    $T^{*}_{ijkl}$. This tensor is not unique: the coarse-graining
    procedure determines it only up to a gauge transformation acting on the
    internal bonds.

    A useful %phase
    indicator for the phases of the system must therefore be built
    from gauge-invariant combinations of $T^{*}$. We use the scalar
    \begin{equation}
      \chi_{\mathrm{inv}}
      \;=\;
      \frac{\bigl(\sum_{ij} T_{ijij}\bigr)^{2}}
           {\sum_{ijkl} T_{ijkl}\,T_{klij}},
      \label{eq:chi_def}
    \end{equation}
    see Refs.~\cite{Gu2009,Yang2017,Guo2024,Homma2024}.
    It is convenient to reshape the four-leg tensor into a matrix by grouping
    opposite pairs of legs,
    %
    %\begin{equation}
      $M_{(ab),(cd)} \equiv T_{abcd}$.
      %\label{eq:reshape}
    %\end{equation}
    %
    Heuristically, the matrix $M$ plays the role of the row-to-row transfer matrix of the
    underlying classical system.
   
    With the reshaping, %\eqref{eq:reshape},
    the two contractions in
    $\chi_{\mathrm{inv}}$ become ordinary matrix traces,

    so that
    \begin{equation}
      %\boxed{\;
      \chi_{\mathrm{inv}}
      = \frac{\bigl(\operatorname{Tr} M\bigr)^{2}}{\operatorname{Tr} M^{2}}
      %\;}
      \label{eq:chi_trace}
    \end{equation}
  
    %\paragraph{Spectral interpretation.}
    The gauge invariant can be interpreted as the inverse participation ratio of the transfer spectrum: it
    counts the effective number of dominant eigenvalues. If $\tilde\lambda_\alpha = \lambda_\alpha / \sum_\beta \lambda_\beta$ with $\{\lambda_\alpha\}$ the eigenvalues of $M$, then

    \begin{equation}
      \chi_{\mathrm{inv}}
      = \frac{\bigl(\sum_\alpha \lambda_\alpha\bigr)^{2}}
             {\sum_\alpha \lambda_\alpha^{2}}
      = \frac{1}{\sum_\alpha \tilde\lambda_\alpha^{2}}.
    \end{equation}

    If the
    leading eigenvalues are degenerate --- as enforced by symmetry at a
    symmetry-broken fixed point --- $\chi_{\mathrm{inv}}$ is an
    integer counting that degeneracy.
    In particular, in the disordered phase, $T^{*}$ factorizes into a  rank-one tensor so that $M$ has a single nonzero eigenvalue and $\chi_\mathrm{inv}=1$.
    More generally, in a phase with a $\mathbb{Z}_p$ symmetry the fully ordered phase supports $p$
    degenerate pure states and $\chi_{\mathrm{inv}}= p$, so the integer value
    of $\chi_{\mathrm{inv}}$ at the fixed point directly reports the number of
    extremal Gibbs states and hence serves as a discrete order parameter for
    the broken symmetry.

    \subsection{Supplementary curves in tricritical region}
    \label{app:tricritial_sm}
    We show some curves similar to Fig.~\ref{fig:tricritical}a for different values of temperature. This highlights the argument provided in the main text: for $T < T_t$, the peak in the gauge invariant is scale invariant, and becomes scale dependent in the tricritical range, up to some value of $T$ at which the peak disappears, see Fig.~\ref{fig:SM_tricritical}.


\begin{thebibliography}{78}%
\makeatletter
\providecommand \@ifxundefined [1]{%
 \@ifx{#1\undefined}
}%
\providecommand \@ifnum [1]{%
 \ifnum #1\expandafter \@firstoftwo
 \else \expandafter \@secondoftwo
 \fi
}%
\providecommand \@ifx [1]{%
 \ifx #1\expandafter \@firstoftwo
 \else \expandafter \@secondoftwo
 \fi
}%
\providecommand \natexlab [1]{#1}%
\providecommand \enquote  [1]{``#1''}%
\providecommand \bibnamefont  [1]{#1}%
\providecommand \bibfnamefont [1]{#1}%
\providecommand \citenamefont [1]{#1}%
\providecommand \href@noop [0]{\@secondoftwo}%
\providecommand \href [0]{\begingroup \@sanitize@url \@href}%
\providecommand \@href[1]{\@@startlink{#1}\@@href}%
\providecommand \@@href[1]{\endgroup#1\@@endlink}%
\providecommand \@sanitize@url [0]{\catcode `\\12\catcode `\$12\catcode
  `\&12\catcode `\#12\catcode `\^12\catcode `\_12\catcode `\%12\relax}%
\providecommand \@@startlink[1]{}%
\providecommand \@@endlink[0]{}%
\providecommand \url  [0]{\begingroup\@sanitize@url \@url }%
\providecommand \@url [1]{\endgroup\@href {#1}{\urlprefix }}%
\providecommand \urlprefix  [0]{URL }%
\providecommand \Eprint [0]{\href }%
\providecommand \doibase [0]{http://dx.doi.org/}%
\providecommand \selectlanguage [0]{\@gobble}%
\providecommand \bibinfo  [0]{\@secondoftwo}%
\providecommand \bibfield  [0]{\@secondoftwo}%
\providecommand \translation [1]{[#1]}%
\providecommand \BibitemOpen [0]{}%
\providecommand \bibitemStop [0]{}%
\providecommand \bibitemNoStop [0]{.\EOS\space}%
\providecommand \EOS [0]{\spacefactor3000\relax}%
\providecommand \BibitemShut  [1]{\csname bibitem#1\endcsname}%
\let\auto@bib@innerbib\@empty
%</preamble>
\bibitem [{\citenamefont {Berezinski{\v i}}(1971)}]{Berezinskii1971}%
  \BibitemOpen
  \bibfield  {author} {\bibinfo {author} {\bibfnamefont {V.~L.}\ \bibnamefont
  {Berezinski{\v i}}},\ }\href@noop {} {\bibfield  {journal} {\bibinfo
  {journal} {Soviet Journal of Experimental and Theoretical Physics}\ }\textbf
  {\bibinfo {volume} {32}},\ \bibinfo {pages} {493} (\bibinfo {year}
  {1971})}\BibitemShut {NoStop}%
\bibitem [{\citenamefont {Berezinski{\v i}}(1972)}]{Berezinskii1972}%
  \BibitemOpen
  \bibfield  {author} {\bibinfo {author} {\bibfnamefont {V.~L.}\ \bibnamefont
  {Berezinski{\v i}}},\ }\href@noop {} {\bibfield  {journal} {\bibinfo
  {journal} {Soviet Journal of Experimental and Theoretical Physics}\ }\textbf
  {\bibinfo {volume} {34}},\ \bibinfo {pages} {610} (\bibinfo {year}
  {1972})}\BibitemShut {NoStop}%
\bibitem [{\citenamefont {Kosterlitz}\ and\ \citenamefont
  {Thouless}(1973)}]{Kosterlitz1973}%
  \BibitemOpen
  \bibfield  {author} {\bibinfo {author} {\bibfnamefont {J.~M.}\ \bibnamefont
  {Kosterlitz}}\ and\ \bibinfo {author} {\bibfnamefont {D.~J.}\ \bibnamefont
  {Thouless}},\ }\href {\doibase 10.1088/0022-3719/6/7/010} {\bibfield
  {journal} {\bibinfo  {journal} {Journal of Physics C: Solid State Physics}\
  }\textbf {\bibinfo {volume} {6}},\ \bibinfo {pages} {1181} (\bibinfo {year}
  {1973})}\BibitemShut {NoStop}%
\bibitem [{\citenamefont {Halperin}\ and\ \citenamefont
  {Nelson}(1978)}]{Halperin1978}%
  \BibitemOpen
  \bibfield  {author} {\bibinfo {author} {\bibfnamefont {B.~I.}\ \bibnamefont
  {Halperin}}\ and\ \bibinfo {author} {\bibfnamefont {D.~R.}\ \bibnamefont
  {Nelson}},\ }\href {\doibase 10.1103/physrevlett.41.121} {\bibfield
  {journal} {\bibinfo  {journal} {Physical Review Letters}\ }\textbf {\bibinfo
  {volume} {41}},\ \bibinfo {pages} {121} (\bibinfo {year} {1978})}\BibitemShut
  {NoStop}%
\bibitem [{\citenamefont {Nelson}\ and\ \citenamefont
  {Halperin}(1979)}]{Nelson1979}%
  \BibitemOpen
  \bibfield  {author} {\bibinfo {author} {\bibfnamefont {D.~R.}\ \bibnamefont
  {Nelson}}\ and\ \bibinfo {author} {\bibfnamefont {B.~I.}\ \bibnamefont
  {Halperin}},\ }\href {\doibase 10.1103/physrevb.19.2457} {\bibfield
  {journal} {\bibinfo  {journal} {Physical Review B}\ }\textbf {\bibinfo
  {volume} {19}},\ \bibinfo {pages} {2457} (\bibinfo {year}
  {1979})}\BibitemShut {NoStop}%
\bibitem [{\citenamefont {Young}(1979)}]{Young1979}%
  \BibitemOpen
  \bibfield  {author} {\bibinfo {author} {\bibfnamefont {A.~P.}\ \bibnamefont
  {Young}},\ }\href {\doibase 10.1103/physrevb.19.1855} {\bibfield  {journal}
  {\bibinfo  {journal} {Physical Review B}\ }\textbf {\bibinfo {volume} {19}},\
  \bibinfo {pages} {1855} (\bibinfo {year} {1979})}\BibitemShut {NoStop}%
\bibitem [{\citenamefont {Murray}\ and\ \citenamefont
  {Van~Winkle}(1987)}]{Murray1987}%
  \BibitemOpen
  \bibfield  {author} {\bibinfo {author} {\bibfnamefont {C.~A.}\ \bibnamefont
  {Murray}}\ and\ \bibinfo {author} {\bibfnamefont {D.~H.}\ \bibnamefont
  {Van~Winkle}},\ }\href {\doibase 10.1103/PhysRevLett.58.1200} {\bibfield
  {journal} {\bibinfo  {journal} {Physical Review Letters}\ }\textbf {\bibinfo
  {volume} {58}},\ \bibinfo {pages} {1200} (\bibinfo {year}
  {1987})}\BibitemShut {NoStop}%
\bibitem [{\citenamefont {Zahn}\ \emph {et~al.}(1999)\citenamefont {Zahn},
  \citenamefont {Lenke},\ and\ \citenamefont {Maret}}]{Zanh_1999}%
  \BibitemOpen
  \bibfield  {author} {\bibinfo {author} {\bibfnamefont {K.}~\bibnamefont
  {Zahn}}, \bibinfo {author} {\bibfnamefont {R.}~\bibnamefont {Lenke}}, \ and\
  \bibinfo {author} {\bibfnamefont {G.}~\bibnamefont {Maret}},\ }\href
  {\doibase 10.1103/PhysRevLett.82.2721} {\bibfield  {journal} {\bibinfo
  {journal} {Physical Review Letters}\ }\textbf {\bibinfo {volume} {82}},\
  \bibinfo {pages} {2721} (\bibinfo {year} {1999})}\BibitemShut {NoStop}%
\bibitem [{\citenamefont {Peng}\ \emph {et~al.}(2010)\citenamefont {Peng},
  \citenamefont {Wang}, \citenamefont {Alsayed}, \citenamefont {Yodh},\ and\
  \citenamefont {Han}}]{Peng2010}%
  \BibitemOpen
  \bibfield  {author} {\bibinfo {author} {\bibfnamefont {Y.}~\bibnamefont
  {Peng}}, \bibinfo {author} {\bibfnamefont {Z.}~\bibnamefont {Wang}}, \bibinfo
  {author} {\bibfnamefont {A.~M.}\ \bibnamefont {Alsayed}}, \bibinfo {author}
  {\bibfnamefont {A.~G.}\ \bibnamefont {Yodh}}, \ and\ \bibinfo {author}
  {\bibfnamefont {Y.}~\bibnamefont {Han}},\ }\href {\doibase
  10.1103/PhysRevLett.104.205703} {\bibfield  {journal} {\bibinfo  {journal}
  {Physical Review Letters}\ }\textbf {\bibinfo {volume} {104}},\ \bibinfo
  {pages} {205703} (\bibinfo {year} {2010})}\BibitemShut {NoStop}%
\bibitem [{\citenamefont {Schockmel}\ \emph {et~al.}(2013)\citenamefont
  {Schockmel}, \citenamefont {Mersch}, \citenamefont {Vandewalle},\ and\
  \citenamefont {Lumay}}]{Schockmel2013}%
  \BibitemOpen
  \bibfield  {author} {\bibinfo {author} {\bibfnamefont {J.}~\bibnamefont
  {Schockmel}}, \bibinfo {author} {\bibfnamefont {E.}~\bibnamefont {Mersch}},
  \bibinfo {author} {\bibfnamefont {N.}~\bibnamefont {Vandewalle}}, \ and\
  \bibinfo {author} {\bibfnamefont {G.}~\bibnamefont {Lumay}},\ }\href
  {\doibase 10.1103/PhysRevE.87.062201} {\bibfield  {journal} {\bibinfo
  {journal} {Physical Review E}\ }\textbf {\bibinfo {volume} {87}},\ \bibinfo
  {pages} {062201} (\bibinfo {year} {2013})}\BibitemShut {NoStop}%
\bibitem [{\citenamefont {Deutschl{\"a}nder}\ \emph {et~al.}(2013)\citenamefont
  {Deutschl{\"a}nder}, \citenamefont {Horn}, \citenamefont {L{\"o}wen},
  \citenamefont {Maret},\ and\ \citenamefont {Keim}}]{Deutschlander_2013}%
  \BibitemOpen
  \bibfield  {author} {\bibinfo {author} {\bibfnamefont {S.}~\bibnamefont
  {Deutschl{\"a}nder}}, \bibinfo {author} {\bibfnamefont {T.}~\bibnamefont
  {Horn}}, \bibinfo {author} {\bibfnamefont {H.}~\bibnamefont {L{\"o}wen}},
  \bibinfo {author} {\bibfnamefont {G.}~\bibnamefont {Maret}}, \ and\ \bibinfo
  {author} {\bibfnamefont {P.}~\bibnamefont {Keim}},\ }\href {\doibase
  10.1103/PhysRevLett.111.098301} {\bibfield  {journal} {\bibinfo  {journal}
  {Physical Review Letters}\ }\textbf {\bibinfo {volume} {111}},\ \bibinfo
  {pages} {098301} (\bibinfo {year} {2013})}\BibitemShut {NoStop}%
\bibitem [{\citenamefont {Chen}\ \emph {et~al.}(1995)\citenamefont {Chen},
  \citenamefont {Kaplan},\ and\ \citenamefont {Mostoller}}]{Chen1995}%
  \BibitemOpen
  \bibfield  {author} {\bibinfo {author} {\bibfnamefont {K.}~\bibnamefont
  {Chen}}, \bibinfo {author} {\bibfnamefont {T.}~\bibnamefont {Kaplan}}, \ and\
  \bibinfo {author} {\bibfnamefont {M.}~\bibnamefont {Mostoller}},\ }\href
  {\doibase 10.1103/PhysRevLett.74.4019} {\bibfield  {journal} {\bibinfo
  {journal} {Physical Review Letters}\ }\textbf {\bibinfo {volume} {74}},\
  \bibinfo {pages} {4019} (\bibinfo {year} {1995})}\BibitemShut {NoStop}%
\bibitem [{\citenamefont {Prestipino}\ \emph {et~al.}(2011)\citenamefont
  {Prestipino}, \citenamefont {Saija},\ and\ \citenamefont
  {Giaquinta}}]{Prestipino2011}%
  \BibitemOpen
  \bibfield  {author} {\bibinfo {author} {\bibfnamefont {S.}~\bibnamefont
  {Prestipino}}, \bibinfo {author} {\bibfnamefont {F.}~\bibnamefont {Saija}}, \
  and\ \bibinfo {author} {\bibfnamefont {P.~V.}\ \bibnamefont {Giaquinta}},\
  }\href {\doibase 10.1103/PhysRevLett.106.235701} {\bibfield  {journal}
  {\bibinfo  {journal} {Physical Review Letters}\ }\textbf {\bibinfo {volume}
  {106}},\ \bibinfo {pages} {235701} (\bibinfo {year} {2011})}\BibitemShut
  {NoStop}%
\bibitem [{\citenamefont {Durand}\ and\ \citenamefont
  {Heu}(2019)}]{Durand2019}%
  \BibitemOpen
  \bibfield  {author} {\bibinfo {author} {\bibfnamefont {M.}~\bibnamefont
  {Durand}}\ and\ \bibinfo {author} {\bibfnamefont {J.}~\bibnamefont {Heu}},\
  }\href {\doibase 10.1103/physrevlett.123.188001} {\bibfield  {journal}
  {\bibinfo  {journal} {Physical Review Letters}\ }\textbf {\bibinfo {volume}
  {123}},\ \bibinfo {pages} {188001} (\bibinfo {year} {2019})}\BibitemShut
  {NoStop}%
\bibitem [{\citenamefont {Saito}(1982)}]{Saito1982}%
  \BibitemOpen
  \bibfield  {author} {\bibinfo {author} {\bibfnamefont {Y.}~\bibnamefont
  {Saito}},\ }\href {\doibase 10.1103/physrevb.26.6239} {\bibfield  {journal}
  {\bibinfo  {journal} {Physical Review B}\ }\textbf {\bibinfo {volume} {26}},\
  \bibinfo {pages} {6239} (\bibinfo {year} {1982})}\BibitemShut {NoStop}%
\bibitem [{\citenamefont {Chui}(1983)}]{Chui1983}%
  \BibitemOpen
  \bibfield  {author} {\bibinfo {author} {\bibfnamefont {S.~T.}\ \bibnamefont
  {Chui}},\ }\href {\doibase 10.1103/physrevb.28.178} {\bibfield  {journal}
  {\bibinfo  {journal} {Physical Review B}\ }\textbf {\bibinfo {volume} {28}},\
  \bibinfo {pages} {178} (\bibinfo {year} {1983})}\BibitemShut {NoStop}%
\bibitem [{\citenamefont {Domany}\ \emph {et~al.}(1984)\citenamefont {Domany},
  \citenamefont {Schick},\ and\ \citenamefont {Swendsen}}]{Domany1984}%
  \BibitemOpen
  \bibfield  {author} {\bibinfo {author} {\bibfnamefont {E.}~\bibnamefont
  {Domany}}, \bibinfo {author} {\bibfnamefont {M.}~\bibnamefont {Schick}}, \
  and\ \bibinfo {author} {\bibfnamefont {R.~H.}\ \bibnamefont {Swendsen}},\
  }\href {\doibase 10.1103/PhysRevLett.52.1535} {\bibfield  {journal} {\bibinfo
   {journal} {Physical Review Letters}\ }\textbf {\bibinfo {volume} {52}},\
  \bibinfo {pages} {1535} (\bibinfo {year} {1984})}\BibitemShut {NoStop}%
\bibitem [{\citenamefont {Strandburg}(1988)}]{Strandburg1988}%
  \BibitemOpen
  \bibfield  {author} {\bibinfo {author} {\bibfnamefont {K.~J.}\ \bibnamefont
  {Strandburg}},\ }\href {\doibase 10.1103/RevModPhys.60.161} {\bibfield
  {journal} {\bibinfo  {journal} {Reviews of Modern Physics}\ }\textbf
  {\bibinfo {volume} {60}},\ \bibinfo {pages} {161} (\bibinfo {year}
  {1988})}\BibitemShut {NoStop}%
\bibitem [{\citenamefont {Marcus}\ and\ \citenamefont
  {Rice}(1996)}]{Marcus1996}%
  \BibitemOpen
  \bibfield  {author} {\bibinfo {author} {\bibfnamefont {A.~H.}\ \bibnamefont
  {Marcus}}\ and\ \bibinfo {author} {\bibfnamefont {S.~A.}\ \bibnamefont
  {Rice}},\ }\href {\doibase 10.1103/PhysRevLett.77.2577} {\bibfield  {journal}
  {\bibinfo  {journal} {Physical Review Letters}\ }\textbf {\bibinfo {volume}
  {77}},\ \bibinfo {pages} {2577} (\bibinfo {year} {1996})}\BibitemShut
  {NoStop}%
\bibitem [{\citenamefont {Van~Enter}\ and\ \citenamefont
  {Shlosman}(2002)}]{VanEnter2002}%
  \BibitemOpen
  \bibfield  {author} {\bibinfo {author} {\bibfnamefont {A.~C.~D.}\
  \bibnamefont {Van~Enter}}\ and\ \bibinfo {author} {\bibfnamefont {S.~B.}\
  \bibnamefont {Shlosman}},\ }\href {\doibase 10.1103/PhysRevLett.89.285702}
  {\bibfield  {journal} {\bibinfo  {journal} {Physical Review Letters}\
  }\textbf {\bibinfo {volume} {89}},\ \bibinfo {pages} {285702} (\bibinfo
  {year} {2002})}\BibitemShut {NoStop}%
\bibitem [{\citenamefont {Prestipino}\ \emph {et~al.}(2012)\citenamefont
  {Prestipino}, \citenamefont {Saija},\ and\ \citenamefont
  {Giaquinta}}]{Prestipino2012}%
  \BibitemOpen
  \bibfield  {author} {\bibinfo {author} {\bibfnamefont {S.}~\bibnamefont
  {Prestipino}}, \bibinfo {author} {\bibfnamefont {F.}~\bibnamefont {Saija}}, \
  and\ \bibinfo {author} {\bibfnamefont {P.~V.}\ \bibnamefont {Giaquinta}},\
  }\href {\doibase 10.1063/1.4749260} {\bibfield  {journal} {\bibinfo
  {journal} {The Journal of Chemical Physics}\ }\textbf {\bibinfo {volume}
  {137}},\ \bibinfo {pages} {104503} (\bibinfo {year} {2012})}\BibitemShut
  {NoStop}%
\bibitem [{\citenamefont {Dudalov}\ \emph {et~al.}(2014)\citenamefont
  {Dudalov}, \citenamefont {Tsiok}, \citenamefont {Fomin},\ and\ \citenamefont
  {Ryzhov}}]{Dudalov2014}%
  \BibitemOpen
  \bibfield  {author} {\bibinfo {author} {\bibfnamefont {D.~E.}\ \bibnamefont
  {Dudalov}}, \bibinfo {author} {\bibfnamefont {E.~N.}\ \bibnamefont {Tsiok}},
  \bibinfo {author} {\bibfnamefont {{\relax Yu}.~D.}\ \bibnamefont {Fomin}}, \
  and\ \bibinfo {author} {\bibfnamefont {V.~N.}\ \bibnamefont {Ryzhov}},\
  }\href {\doibase 10.1063/1.4896825} {\bibfield  {journal} {\bibinfo
  {journal} {The Journal of Chemical Physics}\ }\textbf {\bibinfo {volume}
  {141}},\ \bibinfo {pages} {18C522} (\bibinfo {year} {2014})}\BibitemShut
  {NoStop}%
\bibitem [{\citenamefont {Du}\ \emph {et~al.}(2017)\citenamefont {Du},
  \citenamefont {Doxastakis}, \citenamefont {Hilou},\ and\ \citenamefont
  {Biswal}}]{Du2017}%
  \BibitemOpen
  \bibfield  {author} {\bibinfo {author} {\bibfnamefont {D.}~\bibnamefont
  {Du}}, \bibinfo {author} {\bibfnamefont {M.}~\bibnamefont {Doxastakis}},
  \bibinfo {author} {\bibfnamefont {E.}~\bibnamefont {Hilou}}, \ and\ \bibinfo
  {author} {\bibfnamefont {S.~L.}\ \bibnamefont {Biswal}},\ }\href {\doibase
  10.1039/c6sm02131j} {\bibfield  {journal} {\bibinfo  {journal} {Soft Matter}\
  }\textbf {\bibinfo {volume} {13}},\ \bibinfo {pages} {1548} (\bibinfo {year}
  {2017})}\BibitemShut {NoStop}%
\bibitem [{\citenamefont {Bernard}\ and\ \citenamefont
  {Krauth}(2011)}]{Bernard_2011}%
  \BibitemOpen
  \bibfield  {author} {\bibinfo {author} {\bibfnamefont {E.~P.}\ \bibnamefont
  {Bernard}}\ and\ \bibinfo {author} {\bibfnamefont {W.}~\bibnamefont
  {Krauth}},\ }\href {\doibase 10.1103/PhysRevLett.107.155704} {\bibfield
  {journal} {\bibinfo  {journal} {Physical Review Letters}\ }\textbf {\bibinfo
  {volume} {107}},\ \bibinfo {pages} {155704} (\bibinfo {year}
  {2011})}\BibitemShut {NoStop}%
\bibitem [{\citenamefont {Engel}\ \emph {et~al.}(2013)\citenamefont {Engel},
  \citenamefont {Anderson}, \citenamefont {Glotzer}, \citenamefont {Isobe},
  \citenamefont {Bernard},\ and\ \citenamefont {Krauth}}]{Engel2013}%
  \BibitemOpen
  \bibfield  {author} {\bibinfo {author} {\bibfnamefont {M.}~\bibnamefont
  {Engel}}, \bibinfo {author} {\bibfnamefont {J.~A.}\ \bibnamefont {Anderson}},
  \bibinfo {author} {\bibfnamefont {S.~C.}\ \bibnamefont {Glotzer}}, \bibinfo
  {author} {\bibfnamefont {M.}~\bibnamefont {Isobe}}, \bibinfo {author}
  {\bibfnamefont {E.~P.}\ \bibnamefont {Bernard}}, \ and\ \bibinfo {author}
  {\bibfnamefont {W.}~\bibnamefont {Krauth}},\ }\href {\doibase
  10.1103/PhysRevE.87.042134} {\bibfield  {journal} {\bibinfo  {journal}
  {Physical Review E}\ }\textbf {\bibinfo {volume} {87}},\ \bibinfo {pages}
  {042134} (\bibinfo {year} {2013})}\BibitemShut {NoStop}%
\bibitem [{\citenamefont {Kapfer}\ and\ \citenamefont
  {Krauth}(2015)}]{Kapfer_2015}%
  \BibitemOpen
  \bibfield  {author} {\bibinfo {author} {\bibfnamefont {S.~C.}\ \bibnamefont
  {Kapfer}}\ and\ \bibinfo {author} {\bibfnamefont {W.}~\bibnamefont
  {Krauth}},\ }\href {\doibase 10.1103/PhysRevLett.114.035702} {\bibfield
  {journal} {\bibinfo  {journal} {Physical Review Letters}\ }\textbf {\bibinfo
  {volume} {114}},\ \bibinfo {pages} {035702} (\bibinfo {year}
  {2015})}\BibitemShut {NoStop}%
\bibitem [{\citenamefont {Qi}\ \emph {et~al.}(2014)\citenamefont {Qi},
  \citenamefont {Gantapara},\ and\ \citenamefont {Dijkstra}}]{Qi2014}%
  \BibitemOpen
  \bibfield  {author} {\bibinfo {author} {\bibfnamefont {W.}~\bibnamefont
  {Qi}}, \bibinfo {author} {\bibfnamefont {A.~P.}\ \bibnamefont {Gantapara}}, \
  and\ \bibinfo {author} {\bibfnamefont {M.}~\bibnamefont {Dijkstra}},\ }\href
  {\doibase 10.1039/C4SM00125G} {\bibfield  {journal} {\bibinfo  {journal}
  {Soft Matter}\ }\textbf {\bibinfo {volume} {10}},\ \bibinfo {pages} {5449}
  (\bibinfo {year} {2014})}\BibitemShut {NoStop}%
\bibitem [{\citenamefont {Hajibabaei}\ and\ \citenamefont
  {Kim}(2019)}]{Hajibabaei2019}%
  \BibitemOpen
  \bibfield  {author} {\bibinfo {author} {\bibfnamefont {A.}~\bibnamefont
  {Hajibabaei}}\ and\ \bibinfo {author} {\bibfnamefont {K.~S.}\ \bibnamefont
  {Kim}},\ }\href {\doibase 10.1103/PhysRevE.99.022145} {\bibfield  {journal}
  {\bibinfo  {journal} {Phys. Rev. E}\ }\textbf {\bibinfo {volume} {99}},\
  \bibinfo {pages} {022145} (\bibinfo {year} {2019})}\BibitemShut {NoStop}%
\bibitem [{\citenamefont {Tsiok}\ \emph {et~al.}(2022)\citenamefont {Tsiok},
  \citenamefont {Fomin}, \citenamefont {Gaiduk}, \citenamefont {Tareyeva},
  \citenamefont {Ryzhov}, \citenamefont {Libet}, \citenamefont {Dmitryuk},
  \citenamefont {Kryuchkov},\ and\ \citenamefont {Yurchenko}}]{Tsiok2022}%
  \BibitemOpen
  \bibfield  {author} {\bibinfo {author} {\bibfnamefont {E.~N.}\ \bibnamefont
  {Tsiok}}, \bibinfo {author} {\bibfnamefont {Y.~D.}\ \bibnamefont {Fomin}},
  \bibinfo {author} {\bibfnamefont {E.~A.}\ \bibnamefont {Gaiduk}}, \bibinfo
  {author} {\bibfnamefont {E.~E.}\ \bibnamefont {Tareyeva}}, \bibinfo {author}
  {\bibfnamefont {V.~N.}\ \bibnamefont {Ryzhov}}, \bibinfo {author}
  {\bibfnamefont {P.~A.}\ \bibnamefont {Libet}}, \bibinfo {author}
  {\bibfnamefont {N.~A.}\ \bibnamefont {Dmitryuk}}, \bibinfo {author}
  {\bibfnamefont {N.~P.}\ \bibnamefont {Kryuchkov}}, \ and\ \bibinfo {author}
  {\bibfnamefont {S.~O.}\ \bibnamefont {Yurchenko}},\ }\href {\doibase
  10.1063/5.0075479} {\bibfield  {journal} {\bibinfo  {journal} {The Journal of
  Chemical Physics}\ }\textbf {\bibinfo {volume} {156}},\ \bibinfo {pages}
  {114703} (\bibinfo {year} {2022})}\BibitemShut {NoStop}%
\bibitem [{\citenamefont {Jos{\'e}}\ \emph {et~al.}(1977)\citenamefont
  {Jos{\'e}}, \citenamefont {Kadanoff}, \citenamefont {Kirkpatrick},\ and\
  \citenamefont {Nelson}}]{Jose1977}%
  \BibitemOpen
  \bibfield  {author} {\bibinfo {author} {\bibfnamefont {J.~V.}\ \bibnamefont
  {Jos{\'e}}}, \bibinfo {author} {\bibfnamefont {L.~P.}\ \bibnamefont
  {Kadanoff}}, \bibinfo {author} {\bibfnamefont {S.}~\bibnamefont
  {Kirkpatrick}}, \ and\ \bibinfo {author} {\bibfnamefont {D.~R.}\ \bibnamefont
  {Nelson}},\ }\href {\doibase 10.1103/PhysRevB.16.1217} {\bibfield  {journal}
  {\bibinfo  {journal} {Physical Review B}\ }\textbf {\bibinfo {volume} {16}},\
  \bibinfo {pages} {1217} (\bibinfo {year} {1977})}\BibitemShut {NoStop}%
\bibitem [{\citenamefont {Elitzur}\ \emph {et~al.}(1979)\citenamefont
  {Elitzur}, \citenamefont {Pearson},\ and\ \citenamefont
  {Shigemitsu}}]{Elitzur1979}%
  \BibitemOpen
  \bibfield  {author} {\bibinfo {author} {\bibfnamefont {S.}~\bibnamefont
  {Elitzur}}, \bibinfo {author} {\bibfnamefont {R.~B.}\ \bibnamefont
  {Pearson}}, \ and\ \bibinfo {author} {\bibfnamefont {J.}~\bibnamefont
  {Shigemitsu}},\ }\href {\doibase 10.1103/PhysRevD.19.3698} {\bibfield
  {journal} {\bibinfo  {journal} {Physical Review D}\ }\textbf {\bibinfo
  {volume} {19}},\ \bibinfo {pages} {3698} (\bibinfo {year}
  {1979})}\BibitemShut {NoStop}%
\bibitem [{\citenamefont {Lapilli}\ \emph {et~al.}(2006)\citenamefont
  {Lapilli}, \citenamefont {Pfeifer},\ and\ \citenamefont
  {Wexler}}]{Lapilli2006}%
  \BibitemOpen
  \bibfield  {author} {\bibinfo {author} {\bibfnamefont {C.~M.}\ \bibnamefont
  {Lapilli}}, \bibinfo {author} {\bibfnamefont {P.}~\bibnamefont {Pfeifer}}, \
  and\ \bibinfo {author} {\bibfnamefont {C.}~\bibnamefont {Wexler}},\ }\href
  {\doibase 10.1103/PhysRevLett.96.140603} {\bibfield  {journal} {\bibinfo
  {journal} {Physical Review Letters}\ }\textbf {\bibinfo {volume} {96}},\
  \bibinfo {pages} {140603} (\bibinfo {year} {2006})}\BibitemShut {NoStop}%
\bibitem [{\citenamefont {Li}\ \emph {et~al.}(2022)\citenamefont {Li},
  \citenamefont {Pai},\ and\ \citenamefont {Gu}}]{Li2022}%
  \BibitemOpen
  \bibfield  {author} {\bibinfo {author} {\bibfnamefont {G.}~\bibnamefont
  {Li}}, \bibinfo {author} {\bibfnamefont {K.~H.}\ \bibnamefont {Pai}}, \ and\
  \bibinfo {author} {\bibfnamefont {Z.-C.}\ \bibnamefont {Gu}},\ }\href
  {\doibase 10.1103/PhysRevResearch.4.023159} {\bibfield  {journal} {\bibinfo
  {journal} {Physical Review Research}\ }\textbf {\bibinfo {volume} {4}},\
  \bibinfo {pages} {023159} (\bibinfo {year} {2022})}\BibitemShut {NoStop}%
\bibitem [{\citenamefont {Blume}(1966)}]{Blume1966}%
  \BibitemOpen
  \bibfield  {author} {\bibinfo {author} {\bibfnamefont {M.}~\bibnamefont
  {Blume}},\ }\href {\doibase 10.1103/PhysRev.141.517} {\bibfield  {journal}
  {\bibinfo  {journal} {Physical Review}\ }\textbf {\bibinfo {volume} {141}},\
  \bibinfo {pages} {517} (\bibinfo {year} {1966})}\BibitemShut {NoStop}%
\bibitem [{\citenamefont {Capel}(1966)}]{Capel1966}%
  \BibitemOpen
  \bibfield  {author} {\bibinfo {author} {\bibfnamefont {H.}~\bibnamefont
  {Capel}},\ }\href {\doibase 10.1016/0031-8914(66)90027-9} {\bibfield
  {journal} {\bibinfo  {journal} {Physica}\ }\textbf {\bibinfo {volume} {32}},\
  \bibinfo {pages} {966} (\bibinfo {year} {1966})}\BibitemShut {NoStop}%
\bibitem [{\citenamefont {Blume}\ \emph {et~al.}(1971)\citenamefont {Blume},
  \citenamefont {Emery},\ and\ \citenamefont {Griffiths}}]{Blume1971}%
  \BibitemOpen
  \bibfield  {author} {\bibinfo {author} {\bibfnamefont {M.}~\bibnamefont
  {Blume}}, \bibinfo {author} {\bibfnamefont {V.~J.}\ \bibnamefont {Emery}}, \
  and\ \bibinfo {author} {\bibfnamefont {R.~B.}\ \bibnamefont {Griffiths}},\
  }\href {\doibase 10.1103/physreva.4.1071} {\bibfield  {journal} {\bibinfo
  {journal} {Physical Review A}\ }\textbf {\bibinfo {volume} {4}},\ \bibinfo
  {pages} {1071} (\bibinfo {year} {1971})}\BibitemShut {NoStop}%
\bibitem [{\citenamefont {Kwak}\ \emph {et~al.}(2015)\citenamefont {Kwak},
  \citenamefont {Jeong}, \citenamefont {Lee},\ and\ \citenamefont
  {Kim}}]{Kwak2015}%
  \BibitemOpen
  \bibfield  {author} {\bibinfo {author} {\bibfnamefont {W.}~\bibnamefont
  {Kwak}}, \bibinfo {author} {\bibfnamefont {J.}~\bibnamefont {Jeong}},
  \bibinfo {author} {\bibfnamefont {J.}~\bibnamefont {Lee}}, \ and\ \bibinfo
  {author} {\bibfnamefont {D.-H.}\ \bibnamefont {Kim}},\ }\href {\doibase
  10.1103/PhysRevE.92.022134} {\bibfield  {journal} {\bibinfo  {journal}
  {Physical Review E}\ }\textbf {\bibinfo {volume} {92}},\ \bibinfo {pages}
  {022134} (\bibinfo {year} {2015})}\BibitemShut {NoStop}%
\bibitem [{\citenamefont {Zierenberg}\ \emph {et~al.}(2017)\citenamefont
  {Zierenberg}, \citenamefont {Fytas}, \citenamefont {Weigel}, \citenamefont
  {Janke},\ and\ \citenamefont {Malakis}}]{Zierenberg2017}%
  \BibitemOpen
  \bibfield  {author} {\bibinfo {author} {\bibfnamefont {J.}~\bibnamefont
  {Zierenberg}}, \bibinfo {author} {\bibfnamefont {N.~G.}\ \bibnamefont
  {Fytas}}, \bibinfo {author} {\bibfnamefont {M.}~\bibnamefont {Weigel}},
  \bibinfo {author} {\bibfnamefont {W.}~\bibnamefont {Janke}}, \ and\ \bibinfo
  {author} {\bibfnamefont {A.}~\bibnamefont {Malakis}},\ }\href {\doibase
  10.1140/epjst/e2016-60337-x} {\bibfield  {journal} {\bibinfo  {journal} {The
  European Physical Journal Special Topics}\ }\textbf {\bibinfo {volume}
  {226}},\ \bibinfo {pages} {789} (\bibinfo {year} {2017})}\BibitemShut
  {NoStop}%
\bibitem [{\citenamefont {Moueddene}\ \emph {et~al.}(2024)\citenamefont
  {Moueddene}, \citenamefont {G~Fytas}, \citenamefont {Holovatch},
  \citenamefont {Kenna},\ and\ \citenamefont {Berche}}]{Moueddene2024}%
  \BibitemOpen
  \bibfield  {author} {\bibinfo {author} {\bibfnamefont {L.}~\bibnamefont
  {Moueddene}}, \bibinfo {author} {\bibfnamefont {N.}~\bibnamefont {G~Fytas}},
  \bibinfo {author} {\bibfnamefont {Y.}~\bibnamefont {Holovatch}}, \bibinfo
  {author} {\bibfnamefont {R.}~\bibnamefont {Kenna}}, \ and\ \bibinfo {author}
  {\bibfnamefont {B.}~\bibnamefont {Berche}},\ }\href {\doibase
  10.1088/1742-5468/ad1d60} {\bibfield  {journal} {\bibinfo  {journal} {Journal
  of Statistical Mechanics: Theory and Experiment}\ }\textbf {\bibinfo {volume}
  {2024}},\ \bibinfo {pages} {023206} (\bibinfo {year} {2024})}\BibitemShut
  {NoStop}%
\bibitem [{\citenamefont {Berker}\ and\ \citenamefont
  {Nelson}(1979)}]{Berker1979}%
  \BibitemOpen
  \bibfield  {author} {\bibinfo {author} {\bibfnamefont {A.~N.}\ \bibnamefont
  {Berker}}\ and\ \bibinfo {author} {\bibfnamefont {D.~R.}\ \bibnamefont
  {Nelson}},\ }\href {\doibase 10.1103/PhysRevB.19.2488} {\bibfield  {journal}
  {\bibinfo  {journal} {Phys. Rev. B}\ }\textbf {\bibinfo {volume} {19}},\
  \bibinfo {pages} {2488} (\bibinfo {year} {1979})}\BibitemShut {NoStop}%
\bibitem [{\citenamefont {Cardy}\ and\ \citenamefont
  {Scalapino}(1979)}]{Cardy1979}%
  \BibitemOpen
  \bibfield  {author} {\bibinfo {author} {\bibfnamefont {J.~L.}\ \bibnamefont
  {Cardy}}\ and\ \bibinfo {author} {\bibfnamefont {D.~J.}\ \bibnamefont
  {Scalapino}},\ }\href {\doibase 10.1103/PhysRevB.19.1428} {\bibfield
  {journal} {\bibinfo  {journal} {Phys. Rev. B}\ }\textbf {\bibinfo {volume}
  {19}},\ \bibinfo {pages} {1428} (\bibinfo {year} {1979})}\BibitemShut
  {NoStop}%
\bibitem [{\citenamefont {Dillon}\ \emph {et~al.}(2010)\citenamefont {Dillon},
  \citenamefont {Chiesa},\ and\ \citenamefont {Scalettar}}]{Dillon2010}%
  \BibitemOpen
  \bibfield  {author} {\bibinfo {author} {\bibfnamefont {B.~S.}\ \bibnamefont
  {Dillon}}, \bibinfo {author} {\bibfnamefont {S.}~\bibnamefont {Chiesa}}, \
  and\ \bibinfo {author} {\bibfnamefont {R.~T.}\ \bibnamefont {Scalettar}},\
  }\href {\doibase 10.1103/PhysRevB.82.184421} {\bibfield  {journal} {\bibinfo
  {journal} {Physical Review B}\ }\textbf {\bibinfo {volume} {82}},\ \bibinfo
  {pages} {184421} (\bibinfo {year} {2010})}\BibitemShut {NoStop}%
\bibitem [{\citenamefont {Santos-Filho}\ \emph {et~al.}(2018)\citenamefont
  {Santos-Filho}, \citenamefont {Plascak}, \citenamefont {Sobrinho},\ and\
  \citenamefont {Araujo~Batista}}]{Santos2018}%
  \BibitemOpen
  \bibfield  {author} {\bibinfo {author} {\bibfnamefont {J.~B.}\ \bibnamefont
  {Santos-Filho}}, \bibinfo {author} {\bibfnamefont {J.~A.}\ \bibnamefont
  {Plascak}}, \bibinfo {author} {\bibfnamefont {M.~C.}\ \bibnamefont
  {Sobrinho}}, \ and\ \bibinfo {author} {\bibfnamefont {T.~S.}\ \bibnamefont
  {Araujo~Batista}},\ }\href {\doibase 10.1016/j.physa.2018.03.020} {\bibfield
  {journal} {\bibinfo  {journal} {Physica A: Statistical Mechanics and its
  Applications}\ }\textbf {\bibinfo {volume} {503}},\ \bibinfo {pages} {844}
  (\bibinfo {year} {2018})}\BibitemShut {NoStop}%
\bibitem [{\citenamefont {Skovdal}\ \emph {et~al.}(2023)\citenamefont
  {Skovdal}, \citenamefont {P{\'a}lsson}, \citenamefont {Holdsworth},\ and\
  \citenamefont {Hj{\"o}rvarsson}}]{Skovdal2023}%
  \BibitemOpen
  \bibfield  {author} {\bibinfo {author} {\bibfnamefont {B.~E.}\ \bibnamefont
  {Skovdal}}, \bibinfo {author} {\bibfnamefont {G.~K.}\ \bibnamefont
  {P{\'a}lsson}}, \bibinfo {author} {\bibfnamefont {P.~C.~W.}\ \bibnamefont
  {Holdsworth}}, \ and\ \bibinfo {author} {\bibfnamefont {B.}~\bibnamefont
  {Hj{\"o}rvarsson}},\ }\href {\doibase 10.1103/PhysRevB.107.184409} {\bibfield
   {journal} {\bibinfo  {journal} {Physical Review B}\ }\textbf {\bibinfo
  {volume} {107}},\ \bibinfo {pages} {184409} (\bibinfo {year}
  {2023})}\BibitemShut {NoStop}%
\bibitem [{\citenamefont {Guo}\ and\ \citenamefont {Wei}(2024)}]{Guo2024}%
  \BibitemOpen
  \bibfield  {author} {\bibinfo {author} {\bibfnamefont {W.}~\bibnamefont
  {Guo}}\ and\ \bibinfo {author} {\bibfnamefont {T.-C.}\ \bibnamefont {Wei}},\
  }\href {\doibase 10.1103/PhysRevE.109.034111} {\bibfield  {journal} {\bibinfo
   {journal} {Physical Review E}\ }\textbf {\bibinfo {volume} {109}},\ \bibinfo
  {pages} {034111} (\bibinfo {year} {2024})}\BibitemShut {NoStop}%
\bibitem [{\citenamefont {Homma}\ \emph {et~al.}(2024)\citenamefont {Homma},
  \citenamefont {Okubo},\ and\ \citenamefont {Kawashima}}]{Homma2024}%
  \BibitemOpen
  \bibfield  {author} {\bibinfo {author} {\bibfnamefont {K.}~\bibnamefont
  {Homma}}, \bibinfo {author} {\bibfnamefont {T.}~\bibnamefont {Okubo}}, \ and\
  \bibinfo {author} {\bibfnamefont {N.}~\bibnamefont {Kawashima}},\ }\href
  {\doibase 10.1103/PhysRevResearch.6.043102} {\bibfield  {journal} {\bibinfo
  {journal} {Physical Review Research}\ }\textbf {\bibinfo {volume} {6}},\
  \bibinfo {pages} {043102} (\bibinfo {year} {2024})}\BibitemShut {NoStop}%
\bibitem [{\citenamefont {Swendsen}\ and\ \citenamefont
  {Wang}(1987)}]{Swendsen1987}%
  \BibitemOpen
  \bibfield  {author} {\bibinfo {author} {\bibfnamefont {R.~H.}\ \bibnamefont
  {Swendsen}}\ and\ \bibinfo {author} {\bibfnamefont {J.-S.}\ \bibnamefont
  {Wang}},\ }\href {\doibase 10.1103/PhysRevLett.58.86} {\bibfield  {journal}
  {\bibinfo  {journal} {Physical Review Letters}\ }\textbf {\bibinfo {volume}
  {58}},\ \bibinfo {pages} {86} (\bibinfo {year} {1987})}\BibitemShut {NoStop}%
\bibitem [{\citenamefont {Wolff}(1989)}]{Wolff1989}%
  \BibitemOpen
  \bibfield  {author} {\bibinfo {author} {\bibfnamefont {U.}~\bibnamefont
  {Wolff}},\ }\href {\doibase 10.1103/PhysRevLett.62.361} {\bibfield  {journal}
  {\bibinfo  {journal} {Physical Review Letters}\ }\textbf {\bibinfo {volume}
  {62}},\ \bibinfo {pages} {361} (\bibinfo {year} {1989})}\BibitemShut
  {NoStop}%
\bibitem [{\citenamefont {Levin}\ and\ \citenamefont {Nave}(2007)}]{Levin2007}%
  \BibitemOpen
  \bibfield  {author} {\bibinfo {author} {\bibfnamefont {M.}~\bibnamefont
  {Levin}}\ and\ \bibinfo {author} {\bibfnamefont {C.~P.}\ \bibnamefont
  {Nave}},\ }\href {\doibase 10.1103/PhysRevLett.99.120601} {\bibfield
  {journal} {\bibinfo  {journal} {Physical Review Letters}\ }\textbf {\bibinfo
  {volume} {99}},\ \bibinfo {pages} {120601} (\bibinfo {year}
  {2007})}\BibitemShut {NoStop}%
\bibitem [{\citenamefont {Gu}\ and\ \citenamefont {Wen}(2009)}]{Gu2009}%
  \BibitemOpen
  \bibfield  {author} {\bibinfo {author} {\bibfnamefont {Z.-C.}\ \bibnamefont
  {Gu}}\ and\ \bibinfo {author} {\bibfnamefont {X.-G.}\ \bibnamefont {Wen}},\
  }\href {\doibase 10.1103/PhysRevB.80.155131} {\bibfield  {journal} {\bibinfo
  {journal} {Physical Review B}\ }\textbf {\bibinfo {volume} {80}},\ \bibinfo
  {pages} {155131} (\bibinfo {year} {2009})}\BibitemShut {NoStop}%
\bibitem [{\citenamefont {Yang}\ \emph {et~al.}(2017)\citenamefont {Yang},
  \citenamefont {Gu},\ and\ \citenamefont {Wen}}]{Yang2017}%
  \BibitemOpen
  \bibfield  {author} {\bibinfo {author} {\bibfnamefont {S.}~\bibnamefont
  {Yang}}, \bibinfo {author} {\bibfnamefont {Z.-C.}\ \bibnamefont {Gu}}, \ and\
  \bibinfo {author} {\bibfnamefont {X.-G.}\ \bibnamefont {Wen}},\ }\href
  {\doibase 10.1103/PhysRevLett.118.110504} {\bibfield  {journal} {\bibinfo
  {journal} {Physical Review Letters}\ }\textbf {\bibinfo {volume} {118}},\
  \bibinfo {pages} {110504} (\bibinfo {year} {2017})}\BibitemShut {NoStop}%
\bibitem [{\citenamefont {Portelli}\ \emph {et~al.}(2001)\citenamefont
  {Portelli}, \citenamefont {Holdsworth}, \citenamefont {Sellitto},\ and\
  \citenamefont {Bramwell}}]{Portelli2001}%
  \BibitemOpen
  \bibfield  {author} {\bibinfo {author} {\bibfnamefont {B.}~\bibnamefont
  {Portelli}}, \bibinfo {author} {\bibfnamefont {P.~C.~W.}\ \bibnamefont
  {Holdsworth}}, \bibinfo {author} {\bibfnamefont {M.}~\bibnamefont
  {Sellitto}}, \ and\ \bibinfo {author} {\bibfnamefont {S.~T.}\ \bibnamefont
  {Bramwell}},\ }\href {\doibase 10.1103/PhysRevE.64.036111} {\bibfield
  {journal} {\bibinfo  {journal} {Physical Review E}\ }\textbf {\bibinfo
  {volume} {64}},\ \bibinfo {pages} {036111} (\bibinfo {year}
  {2001})}\BibitemShut {NoStop}%
\bibitem [{\citenamefont {Feynman}(2018)}]{Feynman2018}%
  \BibitemOpen
  \bibfield  {author} {\bibinfo {author} {\bibfnamefont {R.~P.}\ \bibnamefont
  {Feynman}},\ }\href {\doibase 10.1201/9780429493034} {\emph {\bibinfo {title}
  {Statistical {{Mechanics}}: {{A Set Of Lectures}}}}}\ (\bibinfo  {publisher}
  {CRC Press},\ \bibinfo {year} {2018})\BibitemShut {NoStop}%
\bibitem [{\citenamefont {Minnhagen}(1987)}]{Minnhagen1987}%
  \BibitemOpen
  \bibfield  {author} {\bibinfo {author} {\bibfnamefont {P.}~\bibnamefont
  {Minnhagen}},\ }\href {\doibase 10.1103/revmodphys.59.1001} {\bibfield
  {journal} {\bibinfo  {journal} {Reviews of Modern Physics}\ }\textbf
  {\bibinfo {volume} {59}},\ \bibinfo {pages} {1001} (\bibinfo {year}
  {1987})}\BibitemShut {NoStop}%
\bibitem [{\citenamefont {Gupta}\ \emph {et~al.}(1988)\citenamefont {Gupta},
  \citenamefont {DeLapp}, \citenamefont {Batrouni}, \citenamefont {Fox},
  \citenamefont {Baillie},\ and\ \citenamefont {Apostolakis}}]{Gupta1988}%
  \BibitemOpen
  \bibfield  {author} {\bibinfo {author} {\bibfnamefont {R.}~\bibnamefont
  {Gupta}}, \bibinfo {author} {\bibfnamefont {J.}~\bibnamefont {DeLapp}},
  \bibinfo {author} {\bibfnamefont {G.~G.}\ \bibnamefont {Batrouni}}, \bibinfo
  {author} {\bibfnamefont {G.~C.}\ \bibnamefont {Fox}}, \bibinfo {author}
  {\bibfnamefont {C.~F.}\ \bibnamefont {Baillie}}, \ and\ \bibinfo {author}
  {\bibfnamefont {J.}~\bibnamefont {Apostolakis}},\ }\href {\doibase
  10.1103/PhysRevLett.61.1996} {\bibfield  {journal} {\bibinfo  {journal}
  {Physical Review Letters}\ }\textbf {\bibinfo {volume} {61}},\ \bibinfo
  {pages} {1996} (\bibinfo {year} {1988})}\BibitemShut {NoStop}%
\bibitem [{\citenamefont {Villain}(1975)}]{Villain1975}%
  \BibitemOpen
  \bibfield  {author} {\bibinfo {author} {\bibfnamefont {J.}~\bibnamefont
  {Villain}},\ }\href {\doibase 10.1051/jphys:01975003606058100} {\bibfield
  {journal} {\bibinfo  {journal} {Journal de Physique}\ }\textbf {\bibinfo
  {volume} {36}},\ \bibinfo {pages} {581} (\bibinfo {year} {1975})}\BibitemShut
  {NoStop}%
\bibitem [{\citenamefont {Hasenbusch}(2008)}]{Hasenbusch_2008}%
  \BibitemOpen
  \bibfield  {author} {\bibinfo {author} {\bibfnamefont {M.}~\bibnamefont
  {Hasenbusch}},\ }\href {\doibase 10.1088/1742-5468/2008/08/P08003} {\bibfield
   {journal} {\bibinfo  {journal} {Journal of Statistical Mechanics: Theory and
  Experiment}\ }\textbf {\bibinfo {volume} {2008}},\ \bibinfo {pages} {P08003}
  (\bibinfo {year} {2008})}\BibitemShut {NoStop}%
\bibitem [{\citenamefont {Chatterjee}\ \emph {et~al.}(2018)\citenamefont
  {Chatterjee}, \citenamefont {Puri},\ and\ \citenamefont
  {Paul}}]{binder_cumulants_clock}%
  \BibitemOpen
  \bibfield  {author} {\bibinfo {author} {\bibfnamefont {S.}~\bibnamefont
  {Chatterjee}}, \bibinfo {author} {\bibfnamefont {S.}~\bibnamefont {Puri}}, \
  and\ \bibinfo {author} {\bibfnamefont {R.}~\bibnamefont {Paul}},\ }\href
  {\doibase 10.1103/PhysRevE.98.032109} {\bibfield  {journal} {\bibinfo
  {journal} {Phys. Rev. E}\ }\textbf {\bibinfo {volume} {98}},\ \bibinfo
  {pages} {032109} (\bibinfo {year} {2018})}\BibitemShut {NoStop}%
\bibitem [{\citenamefont {West}\ \emph {et~al.}(2015)\citenamefont {West},
  \citenamefont {Garcia-Saez},\ and\ \citenamefont
  {Wei}}]{higher_order_moments}%
  \BibitemOpen
  \bibfield  {author} {\bibinfo {author} {\bibfnamefont {C.~G.}\ \bibnamefont
  {West}}, \bibinfo {author} {\bibfnamefont {A.}~\bibnamefont {Garcia-Saez}}, \
  and\ \bibinfo {author} {\bibfnamefont {T.-C.}\ \bibnamefont {Wei}},\ }\href
  {\doibase 10.1103/PhysRevB.92.115103} {\bibfield  {journal} {\bibinfo
  {journal} {Phys. Rev. B}\ }\textbf {\bibinfo {volume} {92}},\ \bibinfo
  {pages} {115103} (\bibinfo {year} {2015})}\BibitemShut {NoStop}%
\bibitem [{\citenamefont {Nakamoto}\ and\ \citenamefont
  {Takeda}(2016)}]{phys_obs}%
  \BibitemOpen
  \bibfield  {author} {\bibinfo {author} {\bibfnamefont {N.}~\bibnamefont
  {Nakamoto}}\ and\ \bibinfo {author} {\bibfnamefont {S.}~\bibnamefont
  {Takeda}},\ }\href {\doibase 10.24517/00011105} {\bibfield  {journal}
  {\bibinfo  {journal} {Sci. Rep. Kanazawa Univ.}\ }\textbf {\bibinfo {volume}
  {60}},\ \bibinfo {pages} {115103} (\bibinfo {year} {2016})}\BibitemShut
  {NoStop}%
\bibitem [{\citenamefont {Bramwell}\ and\ \citenamefont
  {Holdsworth}(1993)}]{Bramwell1993}%
  \BibitemOpen
  \bibfield  {author} {\bibinfo {author} {\bibfnamefont {S.~T.}\ \bibnamefont
  {Bramwell}}\ and\ \bibinfo {author} {\bibfnamefont {P.~C.~W.}\ \bibnamefont
  {Holdsworth}},\ }\href {\doibase 10.1088/0953-8984/5/4/004} {\bibfield
  {journal} {\bibinfo  {journal} {Journal of Physics: Condensed Matter}\
  }\textbf {\bibinfo {volume} {5}},\ \bibinfo {pages} {L53} (\bibinfo {year}
  {1993})}\BibitemShut {NoStop}%
\bibitem [{\citenamefont {Bramwell}\ and\ \citenamefont
  {Holdsworth}(1994)}]{Bramwell1994}%
  \BibitemOpen
  \bibfield  {author} {\bibinfo {author} {\bibfnamefont {S.~T.}\ \bibnamefont
  {Bramwell}}\ and\ \bibinfo {author} {\bibfnamefont {P.~C.~W.}\ \bibnamefont
  {Holdsworth}},\ }\href {\doibase 10.1103/PhysRevB.49.8811} {\bibfield
  {journal} {\bibinfo  {journal} {Physical Review B}\ }\textbf {\bibinfo
  {volume} {49}},\ \bibinfo {pages} {8811} (\bibinfo {year}
  {1994})}\BibitemShut {NoStop}%
\bibitem [{\citenamefont {Savit}(1980)}]{Savit1980}%
  \BibitemOpen
  \bibfield  {author} {\bibinfo {author} {\bibfnamefont {R.}~\bibnamefont
  {Savit}},\ }\href {\doibase 10.1103/RevModPhys.52.453} {\bibfield  {journal}
  {\bibinfo  {journal} {Reviews of Modern Physics}\ }\textbf {\bibinfo {volume}
  {52}},\ \bibinfo {pages} {453} (\bibinfo {year} {1980})}\BibitemShut
  {NoStop}%
\bibitem [{\citenamefont {Ortiz}\ \emph {et~al.}(2012)\citenamefont {Ortiz},
  \citenamefont {Cobanera},\ and\ \citenamefont {Nussinov}}]{Ortiz2012}%
  \BibitemOpen
  \bibfield  {author} {\bibinfo {author} {\bibfnamefont {G.}~\bibnamefont
  {Ortiz}}, \bibinfo {author} {\bibfnamefont {E.}~\bibnamefont {Cobanera}}, \
  and\ \bibinfo {author} {\bibfnamefont {Z.}~\bibnamefont {Nussinov}},\ }\href
  {\doibase https://doi.org/10.1016/j.nuclphysb.2011.09.012} {\bibfield
  {journal} {\bibinfo  {journal} {Nuclear Physics B}\ }\textbf {\bibinfo
  {volume} {854}},\ \bibinfo {pages} {780} (\bibinfo {year}
  {2012})}\BibitemShut {NoStop}%
\bibitem [{\citenamefont {Di~Francesco}\ \emph {et~al.}(1997)\citenamefont
  {Di~Francesco}, \citenamefont {Mathieu},\ and\ \citenamefont
  {S{\'e}n{\'e}chal}}]{DiFrancesco1997}%
  \BibitemOpen
  \bibfield  {author} {\bibinfo {author} {\bibfnamefont {P.}~\bibnamefont
  {Di~Francesco}}, \bibinfo {author} {\bibfnamefont {P.}~\bibnamefont
  {Mathieu}}, \ and\ \bibinfo {author} {\bibfnamefont {D.}~\bibnamefont
  {S{\'e}n{\'e}chal}},\ }\href {\doibase 10.1007/978-1-4612-2256-9} {\emph
  {\bibinfo {title} {Conformal {{Field Theory}}}}},\ Graduate {{Texts}} in
  {{Contemporary Physics}}\ (\bibinfo  {publisher} {Springer New York},\
  \bibinfo {year} {1997})\BibitemShut {NoStop}%
\bibitem [{\citenamefont {Vanderzande}(1988)}]{Vanderzande1988}%
  \BibitemOpen
  \bibfield  {author} {\bibinfo {author} {\bibfnamefont {C.}~\bibnamefont
  {Vanderzande}},\ }\href {\doibase 10.1103/PhysRevB.38.2865} {\bibfield
  {journal} {\bibinfo  {journal} {Physical Review B}\ }\textbf {\bibinfo
  {volume} {38}},\ \bibinfo {pages} {2865} (\bibinfo {year}
  {1988})}\BibitemShut {NoStop}%
\bibitem [{\citenamefont {Plascak}\ \emph {et~al.}(1993)\citenamefont
  {Plascak}, \citenamefont {Moreira},\ and\ \citenamefont
  {sáBarreto}}]{Plascak1993}%
  \BibitemOpen
  \bibfield  {author} {\bibinfo {author} {\bibfnamefont {J.}~\bibnamefont
  {Plascak}}, \bibinfo {author} {\bibfnamefont {J.}~\bibnamefont {Moreira}}, \
  and\ \bibinfo {author} {\bibfnamefont {F.}~\bibnamefont {sáBarreto}},\
  }\href {\doibase 10.1016/0375-9601(93)90250-4} {\bibfield  {journal}
  {\bibinfo  {journal} {Physics Letters A}\ }\textbf {\bibinfo {volume}
  {173}},\ \bibinfo {pages} {360 } (\bibinfo {year} {1993})}\BibitemShut
  {NoStop}%
\bibitem [{\citenamefont {Taufour}\ \emph {et~al.}(2016)\citenamefont
  {Taufour}, \citenamefont {Kaluarachchi},\ and\ \citenamefont
  {Kogan}}]{Taufour2016}%
  \BibitemOpen
  \bibfield  {author} {\bibinfo {author} {\bibfnamefont {V.}~\bibnamefont
  {Taufour}}, \bibinfo {author} {\bibfnamefont {U.~S.}\ \bibnamefont
  {Kaluarachchi}}, \ and\ \bibinfo {author} {\bibfnamefont {V.~G.}\
  \bibnamefont {Kogan}},\ }\href {\doibase 10.1103/PhysRevB.94.060410}
  {\bibfield  {journal} {\bibinfo  {journal} {Phys. Rev. B}\ }\textbf {\bibinfo
  {volume} {94}},\ \bibinfo {pages} {060410} (\bibinfo {year}
  {2016})}\BibitemShut {NoStop}%
\bibitem [{\citenamefont {Raban}\ \emph {et~al.}(2019)\citenamefont {Raban},
  \citenamefont {Suen}, \citenamefont {Berthier},\ and\ \citenamefont
  {Holdsworth}}]{Raban2019}%
  \BibitemOpen
  \bibfield  {author} {\bibinfo {author} {\bibfnamefont {V.}~\bibnamefont
  {Raban}}, \bibinfo {author} {\bibfnamefont {C.~T.}\ \bibnamefont {Suen}},
  \bibinfo {author} {\bibfnamefont {L.}~\bibnamefont {Berthier}}, \ and\
  \bibinfo {author} {\bibfnamefont {P.~C.~W.}\ \bibnamefont {Holdsworth}},\
  }\href {\doibase 10.1103/PhysRevB.99.224425} {\bibfield  {journal} {\bibinfo
  {journal} {Physical Review B}\ }\textbf {\bibinfo {volume} {99}},\ \bibinfo
  {pages} {224425} (\bibinfo {year} {2019})}\BibitemShut {NoStop}%
\bibitem [{\citenamefont {Colella}\ and\ \citenamefont
  {Suter}(1986)}]{Colella1986}%
  \BibitemOpen
  \bibfield  {author} {\bibinfo {author} {\bibfnamefont {N.~J.}\ \bibnamefont
  {Colella}}\ and\ \bibinfo {author} {\bibfnamefont {R.~M.}\ \bibnamefont
  {Suter}},\ }\href {\doibase 10.1103/PhysRevB.34.2052} {\bibfield  {journal}
  {\bibinfo  {journal} {Phys. Rev. B}\ }\textbf {\bibinfo {volume} {34}},\
  \bibinfo {pages} {2052(R)} (\bibinfo {year} {1986})}\BibitemShut {NoStop}%
\bibitem [{\citenamefont {Toledano}\ \emph {et~al.}(2021)\citenamefont
  {Toledano}, \citenamefont {Pancorbo}, \citenamefont {Alvarellos},\ and\
  \citenamefont {G\'alvez}}]{Toledano2021}%
  \BibitemOpen
  \bibfield  {author} {\bibinfo {author} {\bibfnamefont {O.}~\bibnamefont
  {Toledano}}, \bibinfo {author} {\bibfnamefont {M.}~\bibnamefont {Pancorbo}},
  \bibinfo {author} {\bibfnamefont {J.~E.}\ \bibnamefont {Alvarellos}}, \ and\
  \bibinfo {author} {\bibfnamefont {O.}~\bibnamefont {G\'alvez}},\ }\href
  {\doibase 10.1103/PhysRevB.103.094107} {\bibfield  {journal} {\bibinfo
  {journal} {Phys. Rev. B}\ }\textbf {\bibinfo {volume} {103}},\ \bibinfo
  {pages} {094107} (\bibinfo {year} {2021})}\BibitemShut {NoStop}%
\bibitem [{\citenamefont {Barrat}\ and\ \citenamefont
  {Hansen}(2003)}]{BarratBook2003}%
  \BibitemOpen
  \bibfield  {author} {\bibinfo {author} {\bibfnamefont {J.-L.}\ \bibnamefont
  {Barrat}}\ and\ \bibinfo {author} {\bibfnamefont {J.-P.}\ \bibnamefont
  {Hansen}},\ }\href@noop {} {\emph {\bibinfo {title} {Basic concepts for
  simple and complex liquids}}}\ (\bibinfo  {publisher} {Cambridge University
  Press},\ \bibinfo {year} {2003})\BibitemShut {NoStop}%
\bibitem [{\citenamefont {Bladon}\ and\ \citenamefont
  {Frenkel}(1995)}]{Bladon1995}%
  \BibitemOpen
  \bibfield  {author} {\bibinfo {author} {\bibfnamefont {P.}~\bibnamefont
  {Bladon}}\ and\ \bibinfo {author} {\bibfnamefont {D.}~\bibnamefont
  {Frenkel}},\ }\href {\doibase 10.1103/PhysRevLett.74.2519} {\bibfield
  {journal} {\bibinfo  {journal} {Phys. Rev. Lett.}\ }\textbf {\bibinfo
  {volume} {74}},\ \bibinfo {pages} {2519} (\bibinfo {year}
  {1995})}\BibitemShut {NoStop}%
\bibitem [{\citenamefont {Cheng}\ \emph {et~al.}(1988)\citenamefont {Cheng},
  \citenamefont {Ho}, \citenamefont {Hui},\ and\ \citenamefont
  {Pindak}}]{Cheng1988}%
  \BibitemOpen
  \bibfield  {author} {\bibinfo {author} {\bibfnamefont {M.}~\bibnamefont
  {Cheng}}, \bibinfo {author} {\bibfnamefont {J.~T.}\ \bibnamefont {Ho}},
  \bibinfo {author} {\bibfnamefont {S.~W.}\ \bibnamefont {Hui}}, \ and\
  \bibinfo {author} {\bibfnamefont {R.}~\bibnamefont {Pindak}},\ }\href
  {\doibase 10.1103/PhysRevLett.61.550} {\bibfield  {journal} {\bibinfo
  {journal} {Phys. Rev. Lett.}\ }\textbf {\bibinfo {volume} {61}},\ \bibinfo
  {pages} {550} (\bibinfo {year} {1988})}\BibitemShut {NoStop}%
\bibitem [{\citenamefont {Han}\ \emph {et~al.}(2008)\citenamefont {Han},
  \citenamefont {Ha}, \citenamefont {Alsayed},\ and\ \citenamefont
  {Yodh}}]{Han2008}%
  \BibitemOpen
  \bibfield  {author} {\bibinfo {author} {\bibfnamefont {Y.}~\bibnamefont
  {Han}}, \bibinfo {author} {\bibfnamefont {N.~Y.}\ \bibnamefont {Ha}},
  \bibinfo {author} {\bibfnamefont {A.~M.}\ \bibnamefont {Alsayed}}, \ and\
  \bibinfo {author} {\bibfnamefont {A.~G.}\ \bibnamefont {Yodh}},\ }\href
  {\doibase 10.1103/PhysRevE.77.041406} {\bibfield  {journal} {\bibinfo
  {journal} {Phys. Rev. E}\ }\textbf {\bibinfo {volume} {77}},\ \bibinfo
  {pages} {041406} (\bibinfo {year} {2008})}\BibitemShut {NoStop}%
\bibitem [{\citenamefont {Frenkel}\ and\ \citenamefont
  {Smit}(2023)}]{FrenkelBook2023}%
  \BibitemOpen
  \bibfield  {author} {\bibinfo {author} {\bibfnamefont {D.}~\bibnamefont
  {Frenkel}}\ and\ \bibinfo {author} {\bibfnamefont {B.}~\bibnamefont {Smit}},\
  }\href@noop {} {\emph {\bibinfo {title} {Understanding molecular simulation:
  from algorithms to applications}}}\ (\bibinfo  {publisher} {elsevier},\
  \bibinfo {year} {2023})\BibitemShut {NoStop}%
\bibitem [{\citenamefont {Morita}\ and\ \citenamefont
  {Kawashima}(2019)}]{MORITA201965}%
  \BibitemOpen
  \bibfield  {author} {\bibinfo {author} {\bibfnamefont {S.}~\bibnamefont
  {Morita}}\ and\ \bibinfo {author} {\bibfnamefont {N.}~\bibnamefont
  {Kawashima}},\ }\href {\doibase 10.1016/j.cpc.2018.10.014} {\bibfield
  {journal} {\bibinfo  {journal} {Computer Physics Communications}\ }\textbf
  {\bibinfo {volume} {236}},\ \bibinfo {pages} {65} (\bibinfo {year}
  {2019})}\BibitemShut {NoStop}%
\bibitem [{\citenamefont {Evenbly}\ and\ \citenamefont
  {Vidal}(2015)}]{evenbly2015}%
  \BibitemOpen
  \bibfield  {author} {\bibinfo {author} {\bibfnamefont {G.}~\bibnamefont
  {Evenbly}}\ and\ \bibinfo {author} {\bibfnamefont {G.}~\bibnamefont
  {Vidal}},\ }\href {\doibase 10.1103/PhysRevLett.115.180405} {\bibfield
  {journal} {\bibinfo  {journal} {Phys. Rev. Lett.}\ }\textbf {\bibinfo
  {volume} {115}},\ \bibinfo {pages} {180405} (\bibinfo {year}
  {2015})}\BibitemShut {NoStop}%
\end{thebibliography}
\end{document}